 \RequirePackage{booktabs}
\documentclass[sn-mathphys-ay]{sn-jnl}% Math and Physical Sciences Author Year Reference Style
\usepackage{graphicx}%
\usepackage{multirow}%
\usepackage{amsmath,amssymb,amsfonts}%
\usepackage{amsthm}%
\usepackage{mathrsfs}%
\usepackage[title]{appendix}%
\usepackage{xcolor}%
\usepackage{textcomp}%
\usepackage{manyfoot}%
\usepackage{algorithm}%
\usepackage{algorithmicx}%
\usepackage{algpseudocode}%
\usepackage{listings}%

\usepackage{caption}
\usepackage{subcaption}
\usepackage{multirow}
\usepackage{longtable}
\usepackage{hyperref}
\theoremstyle{thmstyleone}%
\theoremstyle{thmstyletwo}%

\theoremstyle{thmstylethree}%

\begin{document}

\title{Agent-Based Modelling of Older Adult Needs for Autonomous Mobility-on-Demand: A Case Study in Winnipeg, Canada}


\author[1,2]{\fnm{Manon} \sur{Prédhumeau}}\email{m.predhumeau@leeds.ac.uk}
\author[1]{\fnm{Ed} \sur{Manley}}\email{e.j.manley@leeds.ac.uk}

\affil*[1]{\orgdiv{School of Geography}, \orgname{University of Leeds}, \orgaddress{\city{Leeds}, \country{UK}}}

\affil[2]{\orgdiv{UMR 5505}, \orgname{IRIT, Université Toulouse Capitole}, \orgaddress{\city{Toulouse}, \country{France}}}


\abstract{As the populations continue to age across many nations, ensuring accessible and efficient transportation options for older adults has become an increasingly important concern. Autonomous Mobility-on-Demand (AMoD) systems have emerged as a potential solution to address the needs faced by older adults in their daily mobility. However, estimation of older adult mobility needs, and how they vary over space and time, is crucial for effective planning and implementation of such service, and conventional four-step approaches lack the granularity to fully account for these needs. To address this challenge, we propose an agent-based model of older adults mobility demand in Winnipeg, Canada. The model is built for 2022 using primarily open data, and is implemented in the Multi-Agent Transport Simulation (MATSim) toolkit. After calibration to accurately reproduce observed travel behaviors, a new AMoD service is tested in simulation and its potential adoption among Winnipeg older adults is explored. The model can help policy makers to estimate the needs of the elderly populations for door-to-door transportation and can guide the design of AMoD transport systems.}




\maketitle

\section{Introduction}
% State the objectives of the work and provide an adequate background, avoiding a detailed literature survey or a summary of the results.

%Context
According to the United Nations, the share of the global population aged 65 years or over is expected to increase from 9.1\% in 2019 to around 11.7 \% in 2030 \citep{un_2019}.
Canada is among the countries with a particularly high proportion of elderly people. As of 2019, 17.6\% of Canada’s population are older adults (i.e., aged 65 or over) and this proportion is expected to reach 22.8\% by 2030 \citep{un_2019}. Such demographic trends pose challenges to the existing transportation infrastructure and services, as it requires ensuring accessible and efficient transport options for the increasing number of older adults with diverse mobility needs. Indeed, the ageing population is characterised by diverse transportation access and mobility requirements, including social engagements, medical appointments, grocery shopping, and recreational activities \citep{haustein_mobility_2012}. However, physical limitations, cognitive decline, and reduced access to private vehicles often restrict older adults' ability to independently travel and participate fully in society \citep{luiu_travel_2021}. Consequently, there is a growing demand for flexible transportation services that can adapt to the needs of older adults \citep{faber_how_2020}.

%Purpose/importance
To address this challenge, a rising trend is to move towards demand-responsive transportation with shared electric autonomous vehicles, sometimes referred to as ``shared autonomous vehicles” (shared AV), “demand-responsive transportation” (DRT) or “autonomous mobility-on-demand” (AMoD). AMoD provides on-demand, door-to-door service with autonomous vehicles, based on real-time users requests. It allows for flexible routing, scheduling, and vehicle allocation, enabling more efficient use of resources. This new mean of transportation could present a sustainable and affordable alternative to private cars for older adults \citep{kersting_2021}, while the on-demand door-to-door aspect maintains the flexibility and independence provided by a private transport. However, previous implementations have shown that DRT can be very failure prone; 50\% last less than 7 years \citep{currie_why_2020}. To be effective an AMoD system must be carefully planned, and fully address the users' diverse and complex needs. In particular, estimating the potential demand is a critical step before local transport authorities introduce a new tailored AMoD service.

% Modelling challenge
Many conventional approaches towards modelling mobility demand are limited by a lack of granularity at critical scales. Traditional four-step approaches, which incrementally predict trip flows, mode uses, and route choices, work on the aggregate zonal scale. While it is possible to model sub-populations separately within these model configurations \citep{schmocker2005estimating}, complex inter-dependencies that might exist between demographic factors, trip generation, and other components (e.g. mode choice) are lost, and planning is unable to account for spatial and temporal heterogeneity in service utility. Both spatial and temporal factors are important to address the need for flexible and responsive services for users with diverse needs and preferences.

%Related work
An emerging alternative to traditional transport modelling approaches is agent-based modelling. These approaches provide greater flexibility by capturing heterogeneity and dependency in and between contextual factors (e.g. demographics, health, household structures) and mobility preferences and behaviours. There are several agent-based modelling platforms in use globally, both with generic and transportation-focused design. As a result of this growth in interest (see \cite{bastarianto2023agent} for a recent review), several research works have explored AMoD as a future transportation mode and its potential adoption by citizens using agent-based modelling (ABM): in Singapore \citep{oh_assessing_2020}, in the United States (Sioux Falls \citep{wang_simulation_2018}, Austin \citep{fagnant_2016}, Wayne County \citep{kagho_2021} and Birmingham \citep{salman_2023}), in Germany (Berlin \citep{ziemke_2019, kaddoura_impact_2020}, Munich \citep{militao_optimal_2021}, Bremerhaven \citep{schluter_impact_2021} and rural town of Colditz \citep{viergutz_demand_2019}), in France (Paris \citep{horl_2019}), in the UK (Bristol \citep{franco_demand_2020}), in Austria (Vienna \citep{muller_2021}) and in Switzerland (Zurich \citep{muller_2014, boesch_2016}).

%Problematic to be solved
However, previous demand models have focused on the general population and commuting mobility. Yet it is the older populations that have the most to benefit from DRT services \citep{kovacs2020aged, zandieh2021mobility}, no research has sought to explicitly model and supply for the demands of older populations. In this paper, we propose a precise spatial model that focus on the heterogeneous mobility needs of older adults, and can serves as a basis to built up an AMoD service for older adults.

%case study and approach
In order to evaluate the older adults needs for AMoD, we focused on Winnipeg, in Canada. Winnipeg is the capital city of the province of Manitoba, and the sixth most populous city in Canada. In 2021 Winnipeg had 749,607 inhabitants with 17 \% of the population aged 65 or older \citep{census_2021}. However, older adults spatial distribution is highly heterogeneous. Depending on the dissemination area considered, older adults can represent up to 70 \% of the population.
The city is connected by a dense road network, and bus services, overseen by the Winnipeg Transit public agency. The ageing population in Winnipeg highlights the need for new age-friendly transportation to ensure a high quality of life for older adults. To address this concern, local authorities are interested in integrating AMoD services dedicated to older adults in their future transportation planning.

To predict the older adults demand for AMoD in Winnipeg, we build on prior work and use an ABM approach, given its ability to represent heterogeneous human behaviours. The model allows to consider the heterogeneity in populations, and to perform large transport simulations at a very fine spatio-temporal scale. 

This work provides the following contributions:
\begin{enumerate}
   \item We designed an ABM of the elderly population daily mobility in Winnipeg. Using open data, we generated a 2022 synthetic population, an activity-based model and a model of the environment including the road network, public transit supply, buildings and facilities. The entire Winnipeg population is modelled - not only older adults - in order to reproduce interactions between individuals in households and in the environment. However, the general population is modelled with less details. We ran simulations with Multi-Agent Transport Simulation (MATSim), an open-source framework for transport simulations, and calibrated the model to accurately reproduce observed travel behaviours.
   
    \item A new AMoD service was introduced and additional simulations were run with this new mobility option for older adults. A simple AMoD supply is tested in simulation to explore the potential adoption of such a service among Winnipeg older adults.
\end{enumerate}

%Outline
The paper is structured as follows. Section \ref{model} explains the ABM and its design process. Section \ref{simulation} describes the MATSim simulations setup and the model's calibration.
Section \ref{validation} presents the model validation against real-world observations.
Section \ref{exploration} tests the implementation of an AMoD service in simulation and presents insights from two demand scenarios and three AMoD fleet sizes. Finally, Section \ref{conclu} presents the main conclusions.

\section{Agent-based modelling of older adult mobility}\label{model}
\begin{figure}[ht!]
    \centering
    \includegraphics[width=0.9\textwidth]{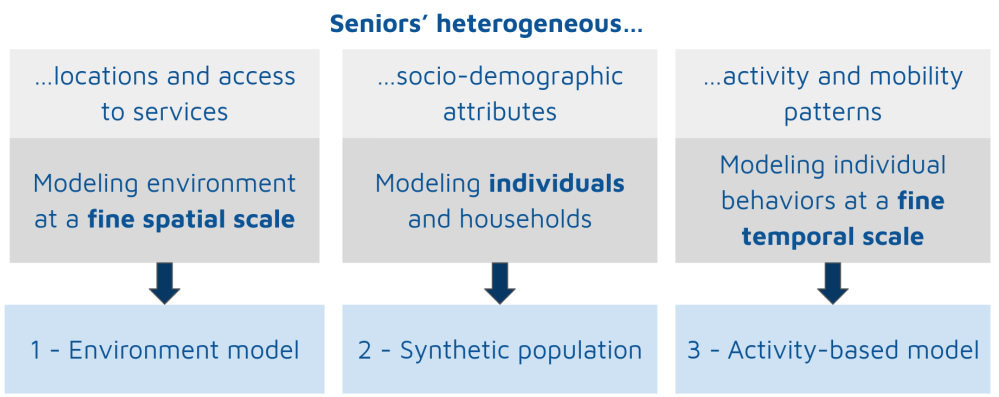}
    \caption{Diagram of the three main components of the proposed ABM: an environment model, a synthetic population, and an activity-based model.}
    \label{fig:fig1}
\end{figure}

In predicting the mobility of older adult populations, factors such as health status, access to transportation options, and proximity to services are known to play an important role \citep{luiu_travel_2021, fatima_elderly_2020}. As such, any prediction of mobility behaviour must consider these factors, as well as inherent interdependencies between factors, heterogeneity over the population, and variation over space and time.
The approach is presented in Figure \ref{fig:fig1}. To account for the heterogeneity in the seniors locations and access to services, we used OpenStreetMap (OSM) \citep{osm} to model the infrastructure of the city of Winnipeg at a fine spatial scale. Due to the diverse sociodemographic attributes of elderly individuals, we used a synthetic population of individuals, i.e. the agents, to model Winnipeg 2022 population.  Finally, in order to capture the variation in their activity and mobility patterns, we developed an activity-based model that considers individual behaviours at a fine temporal scale, i.e. the agents' behaviours.
 The approach is close to the one adopted by \cite{horl_synthetic_2021} to generate a synthetic population and travel demand for Île-de-France. However, our approach is adapted to Canadian open data and includes an additional projection step to produce a synthetic population for future years.

\subsection{Environment model}\label{env_model}
OSM data for Winnipeg was used as a basis to model the road network, the public transit services, and the buildings and facilities. The environment modelling has been described in our previous work \citep{predhumeau_cupum_2023} and is summarised here.

\begin{figure}[ht]
    \centering
    \begin{subfigure}[b]{0.39\textwidth}
         \centering
         \includegraphics[width=\textwidth]{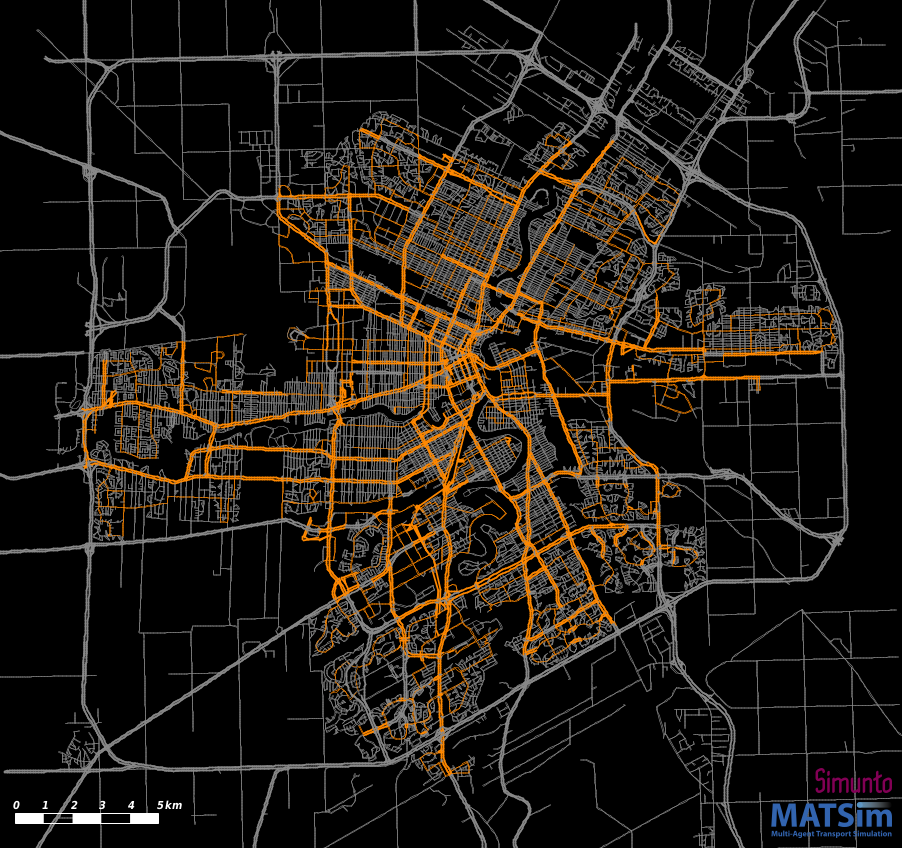}
         \caption{}
         \label{fig:fig3a}
     \end{subfigure}
     \hfill
     \begin{subfigure}[b]{0.42\textwidth}
         \centering
         \includegraphics[width=\textwidth]{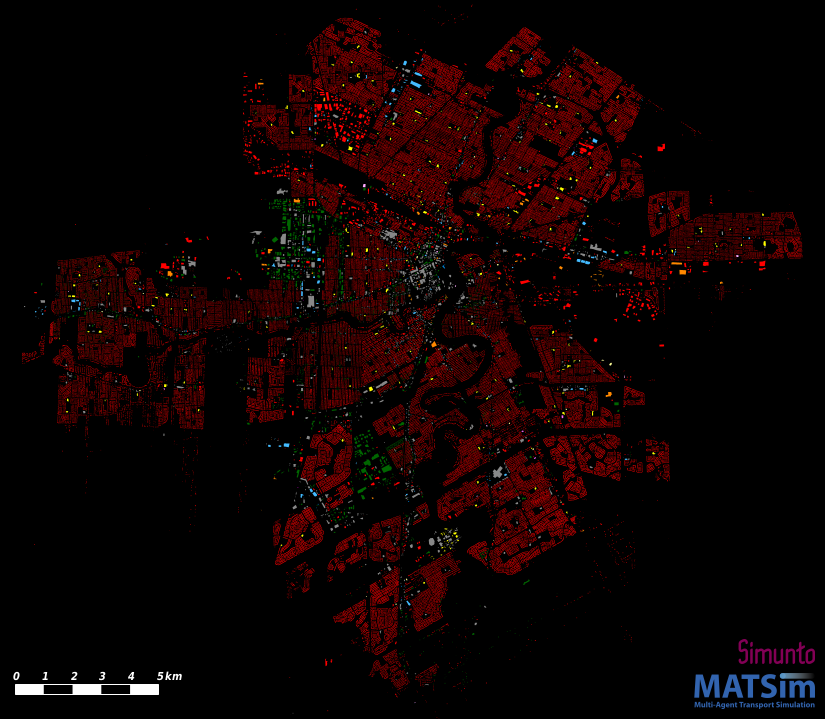}
         \caption{}
         \label{fig:fig3b}
     \end{subfigure}
    \caption{(a) Model of Winnipeg road network in grey and bus routes on a Tuesday service in orange. Road width is proportional to the road capacity. (b) Model of Winnipeg buildings and facilities. Residential buildings are in red, shops and sustenance in blue, education and civic amenities in yellow, sport and entertainment facilities in pink, healthcare buildings in light green, industrial and transport in orange, other land use in dark green and mixed land use in grey. Visualisation realised with Simunto.}
    \label{fig:fig3}
\end{figure}

The Winnipeg road network, and public transit network and service have been modelled using \textit{PT2MATSim} \citep{pt2matsim}, a MATSim plugin. PT2MATSim converts OSM roads into a multimodal network of links, representing the road network and buses lanes, and nodes, representing intersections or connection between links. Winnipeg's road network consists of 97,902 links and 47,352 nodes. Flow capacities for links are assigned from OSM tags and have been checked visually.
PT2MATSim has also been used to convert Winnipeg Transit GTFS feed to a transit schedule and stop sequences for a typical Tuesday (14th of December 2021). The model consists of 5,345 stops served by 84 bus routes, mapped to the multimodal network (Figure \ref{fig:fig3a}).

Buildings and facilities were modelled using three data sources: building polygons from Microsoft Building Footprints \citep{microsoft}, addresses from Winnipeg open data portal \citep{addresses} and facilities from OSM points of interest \citep{osm}. 

Polygons with an area inferior to 30 m$^2$ or over 350,000 m$^2$ have been filtered out in order to keep only the building polygons. The building polygons have been spatially joined with the City of Winnipeg Property Addresses and Coordinates, in order to filter out buildings without addresses.
Finally, OSM building tags and points of interest have been spatially joined with the building polygons. Buildings have been assigned at least one functional type, inferred from the associated OSM data, among industrial, civic amenity, sport facility, sustenance, education, transport, financial, healthcare, entertainment, shop, office or other use (Figure \ref{fig:fig3b}). A building may be multi-types, e.g. a shop on the ground floor and a sport facility on the first floor. Building tagged as residential or with no tag are considered as residential, and are converted into residences by dividing the building area and levels by the average Canadian living area.

\subsection{Synthetic population} \label{syn_pop}
In order to accurately model the Winnipeg population, we created a 2022 synthetic population. The synthetic population design process has been described for 2023 for all Canada by \cite{predhumeau_2023} and \cite{predhumeau_cupum_2023} and is thus briefly summarised here.

The synthetic population was generated for Manitoba province at the dissemination area (DA) level using an hybrid approach called Quasirandom Integer Sampling of Iterative Proportional Fitting \citep{smith_2017}. This approach was applied on the 2016 Canadian Census data \citep{census_2016}. Each individual in the Individual Public Use Microdata File \citep{indiv_pumf} was assigned a weight such that the weighted population at the DA level fits socio-demograpĥic attributes distributions defined by the Census Profile \citep{census_2016}. The 2016 Manitoba synthetic population was then projected to represent the 2022 Manitoba population, using provincial population projections by age and sex \citep{proj_2018}. For each age group and sex, the 2016 synthetic population is resampled (i.e. individuals are duplicated or deleted) to cover the population difference between 2016 and 2022. 
The 2022 Winnipeg synthetic population was extracted from the 2022 Manitoba synthetic population by filtering out the DAs outside Winnipeg. 

A key influence on the modal behaviour of older adults is the household composition; living with a partner can enable access to car as a passenger. The synthetic population therefore includes households.
Each 2022 synthetic individual identified as a primary household `maintainer' was assigned to their own household. The remaining non-maintainer individuals are subsequently assigned to households within their DA to match their household size attribute. To form households, individuals are randomly drawn following the distribution of individuals’ age group and sex by primary maintainer's age group and sex inferred from the 2016 Hierarchical Public Use Microdata File \citep{hh_pumf}.

Besides age, sex and household composition, transportation access and health status have also been identified as important factors of older adults' mobility \citep{luiu_travel_2021}.
Winnipeg individuals' attributes have thus been complemented with additional ``health status'', ``driving license'' and ``residential building'' attributes as described by \cite{predhumeau_cupum_2023}.
A ``health status” attribute has been probabilistically added to synthetic individuals according to the Canadian Community Health Survey 2021 \citep{health_survey}. Statistics on perceived health by age group and sex for Manitoba have been used to assign individuals health status attribute among \{0: poor or fair health; 1: good health; 2: very good or excellent health\}. Similarly, a ``driving licence” attribute has been attributed to individuals according to 2020 statistics from Manitoba Public Insurance \citep{mpi}. Probabilities to own a driving license by age and gender have been derived from the 2020 Traffic Collision Statistics Report which reports licensed drivers by age group, gender and driver status for Manitoba.
The correlations between these two new variables and both age and sex are maintained because the new variables are assigned according to a probability based on age and sex. However, other correlations (between health and income level, for example) are not explicitly maintained  due to the unavailability of such data.
Finally, households have been assigned a residential building within their DA using the buildings model described in Section \ref{env_model}.

\begin{figure}[!ht]
    \centering
    \includegraphics[trim=0cm 0cm 0cm 0cm ,clip, width=0.7\textwidth]{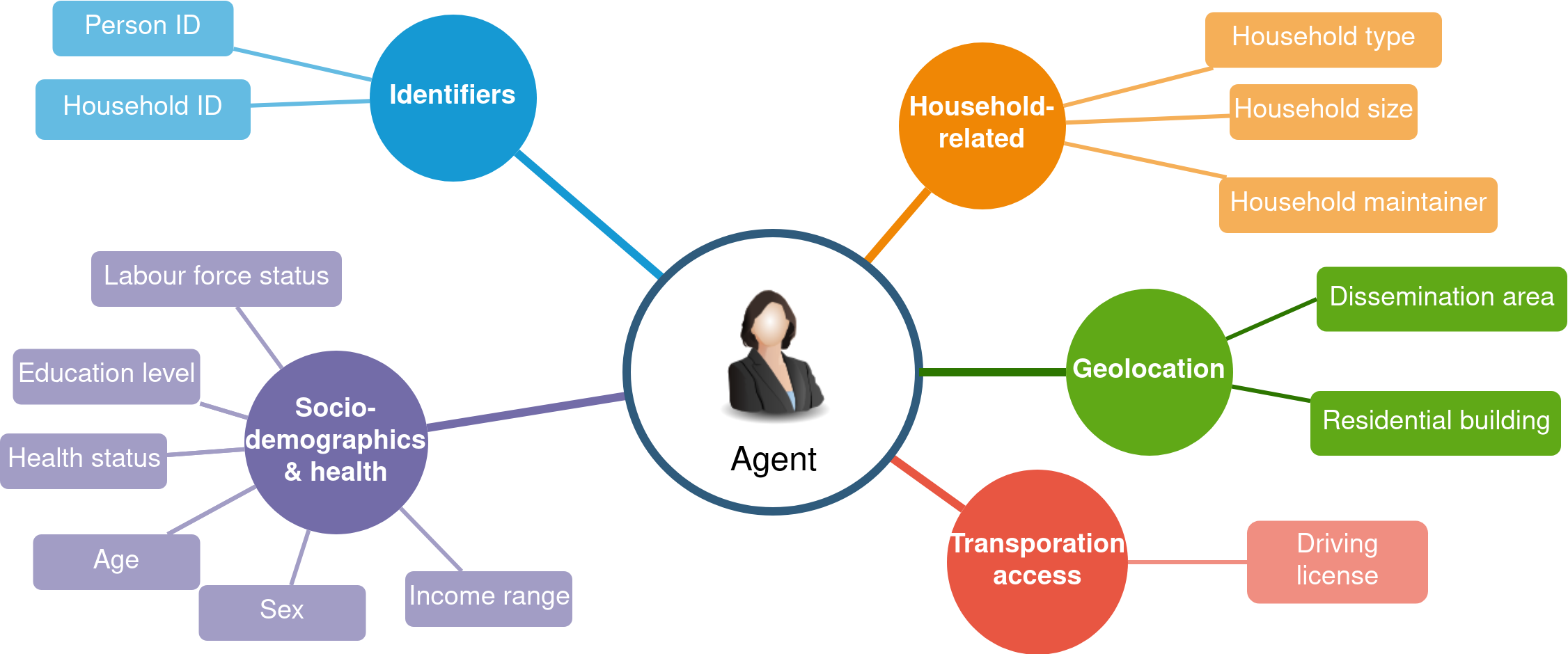}
    \caption{Synthetic population individual attributes.}
    \label{fig:fig2}
\end{figure}

The final 2022 Winnipeg synthetic population is composed of 768,845 individuals (127,314 older adults), with a set of attributes regarding their socio-demographics, household, geolocation, and transportation access (Figure \ref{fig:fig2}).

\subsection{Activity-based model}
\subsubsection{Definition}
\begin{figure}[ht]
     \centering
     \includegraphics[width=\textwidth]{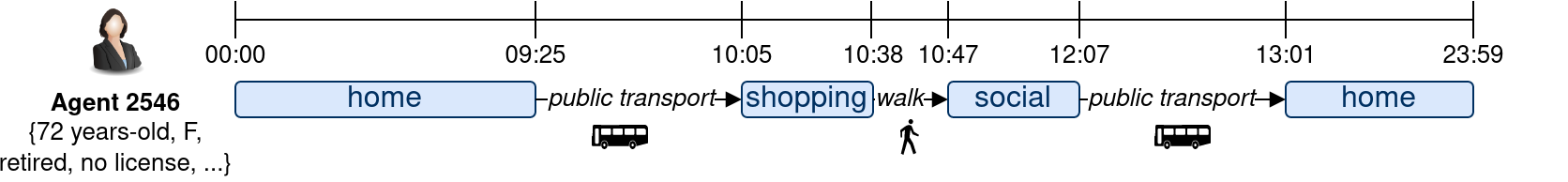}
     \caption{Example of activity schedule for one agent.}
     \label{fig:fig4}
 \end{figure}

In activity-based models, travel decisions are not solely based on the characteristics of the transportation system but are influenced by the purpose and scheduling of people activities. This type of modelling is used in transportation planning to model and simulate individual-level travel behaviour by considering various activities people engage in throughout their daily lives, such as work, school, shopping, and recreational activities. An activity-based model simulates individual activities and trips over a day, at a fine temporal scale e.g. the minute scale, as illustrated on Figure \ref{fig:fig4}.

To build the activity-based model, Canadian older adults most common out-of-home activities have been identified from \cite{newbold_travel_2005} as:
\begin{itemize}
    \item shopping: purchasing consumer goods, products, or service;
    \item socialising: visiting friends or relatives, communicating;
    \item eating out: eating or drinking at a restaurant, bar, club;
    \item entertaining: hobbies or leisure activities like attending cinema, library, museum, theatre, sporting events, etc.;
    \item exercising: recreational or competitive sport, walking or hiking;
    \item medical: health visit;
    \item working: exercising paid work activities;
    \item escorting someone: accompanying to or from school, bus stop, sports, appointments, shopping.
    \item other: volunteering, religious activities, administrative activities, civic participation
\end{itemize}

As all the population is modelled, and not only older adults, the activity-based model also includes an activity type ``education”, which is an important out-of-home activity for kids and young adults. Finally, an activity type ``home” was added, to represent agents being at home.
Similarly, the 5 main transportation modes of older adults \citep{newbold_travel_2005} have been included into the model: car as a driver, car as a passenger (i.e. ride), public transit, walk and bike.

\subsubsection{Statistical matching}
In the absence of a national or a recent local travel survey, the activities generation and scheduling have been guided by the 2015 Time Use Survey (TUS) data \citep{tus_2015}. The TUS is a survey on people time use that contains demographics and daily diaries of 17390 Canadians. It was collected from April 2015 to April 2016, as part of the General Society Survey, and the microdata file containing the respondents data has been released in November 2017. The data consists of a 24h period report (rounded to 5 min) of any activity and travel, by one respondent by household, and some socio-demographic variables such as age group, sex, living arrangement, health status, etc. The survey excluded individuals aged under 15 years and residents of institutions.

Activity and trip characteristics correlate with socio-demographic characteristics; therefore a unconstrained statistical matching approach \citep{namazirad_2017} has been applied. The approach matches each individual in the synthetic population with one of the TUS respondents, based on their common socio-demographic characteristics. Each synthetic individual was matched to its closest TUS record using a ball-tree data structure, i.e. a binary tree for organising points in a multi-dimensional space for efficient nearest neighbour search.

An important weight (arbitrarily fixed at 100 to ease the matching quality analysis) has been applied to attributes that are key drivers of older adults mobility, in order to discourage mismatching these attributes. The older adults age ranges have been assigned a 100 weight because they are the focus of this study and accurately matching them is more important than the rest of the population. The sex attribute has been assigned a 100 weight because gender is one of the main factor influencing older adults mobility. Similarly, in case of poor or fair health, a weight of 100 was used as this is an important obstacle to mobility. The absence of driving license was also assigned a 100 weight as car access is key for older adults mobility and modal choice. Finally, the absence of children in the household and the labour force status have been assign a 100 weight because a mismatch could lead to inconsistent behaviours, e.g. escorting a child to school while there is no child in the household or spending the day at work while not in the labour force.

As rural and urban mobility patterns differ significantly, we removed 4071 TUS respondents living outside large urban population centres (Census Metropolitan Areas or Census Agglomeration) before the matching. Remaining respondents with at least one missing activity location (124), trips by plane (56) or boat (33) or with car trips more than one hour long (1012) have been removed as these trips are not considered in the model. Similarly, it was found that 17\% of the trips might be missing in the remaining records, as sometimes respondents indicated to be at a location and then a different location, with no travel activity between. The respondents with identified missing trips (3644) have been removed from the sample before the matching, as travel characteristics are essential to build the activity-based model. Finally, activity and mobility patterns differ between week days and weekends, and for a sake of consistency the model can not mix week day schedules and weekend schedules. We thus removed the weekend records (2427) and retained 6023 TUS week days respondents for the statistical matching.

The attributes of the synthetic population and the TUS respondents have been aligned in order to perform the matching.
At the end of the matching, 2684 different TUS respondents and daily diaries were matched to the 768845 agents, with 692 daily diaries matched to the 127314 older adult agents. To evaluate the matching quality, the pairs of matched individuals have been compared.

39.9\% of synthetic individuals are perfectly matched regarding all their attributes. 32.6\% of synthetic individuals are matched with a distance of 1 and 24.5\% with a distance between 2 and 5. Finally, 3\% are matched with a distance $>$ 5 and 1.5\% with a distance $\geq$ 100, i.e. a mismatch on an important attribute. When considering synthetic older adults, 50.8\% are perfectly matched, 18.9\% are matched with a distance of 1, and 19.7\% with a distance between 2 and 5. Finally, 10.6\% are matched with a distance $>$ 5 and 7.5\% with a distance $\geq$ 100.

An analysis of the matching quality at the attribute level in Table \ref{tab2} shows that more than 95\% of older adults are perfectly matched regarding sex, age range, labour force status, driving license ownership, and child presence in the household, i.e. the key attributes for older adults mobility that were assigned an important weight for the matching. Similarly, 99.93\% of older adults are well matched regarding the presence or absence of poor or fair health status. For the other attributes, the matching is less precise but still relevant, more than 77\% of agents are perfectly matched for each attribute. The general population matching scores are similar to those of older adults.

\begin{table}[h]
\caption{Statistical matching scores for older adults in bold, and total population in brackets, for each attribute}\label{tab2}
\begin{tabular*}{\textwidth}{@{\extracolsep\fill}lcccccc}
\toprule%
Attribute & \% of agents perfectly matched& \% of agents quasi perfectly matched\\
& & i.e. maximum one class shift\\
\midrule
Sex & \textbf{99.79} (99.87) & - \\
Age range& \textbf{96.57} (84.67) & \textbf{99.34} (96.78)\\ 
Health status & \textbf{88.55} (87.91) & \textbf{99.93} (99.99) \\
 Income range& \textbf{80.69} (84.61) & \textbf{97.44} (98.75)\\
Labour force status & \textbf{99.75 to 99.99} (99.81 to 99.99)&-\\
Qualification & \textbf{84.57} (81.06) &-\\
Driving license & \textbf{99.76} (99.86)&-\\ 
Household size & \textbf{77.87} (80.8)&\textbf{90.72} (95.48)\\
Children's presence & \textbf{98.49} (99.65)&-\\ 
Household type & \textbf{82.69} (92.1)&-\\ 
\botrule
\end{tabular*}
\end{table}

\subsubsection{Schedules generation}\label{acbm}
Each agent was assigned its matched TUS and corresponding daily schedule. The TUS contains for each respondent a daily schedule as a list of episodes with a start time, a duration, a location type (e.g. at home, at place of work or school, grocery store, etc.) and an activity type (e.g. sleeping, eating, paid work, shopping for or buying goods, exercising, etc.). A trip is a travel from one activity to another and is recorded in the TUS as an episode with an activity type ``travel” and a location type corresponding to the travel mode (car, ride, walk, bus, bike, etc.).

For each agent, episodes are converted into a consistent schedule alternating activities and trips between activities.
If there is no trip, the agent stays at home.
MATSim requires agents to start and end the day with the same activity type. If the agent is not home at the start or end of the day, a ``home” activity and the preceding/following trip are added before the first activity of after the last activity.
If consecutive trips identified, they are merged into one trip.

The schedule generation algorithm is as follows:
\begin{enumerate}
   \item For each trip episode \textit{\textbf{T}}, the preceding activity episode \textit{\textbf{A}} and preceding trip \textit{\textbf{T-1}} are retrieved. \textit{\textbf{A}} activity type is identified using a combination of the episode location type and episode activity type.
    
    \item Then a location for \textit{\textbf{A}} is drawn from Winnipeg facilities and residences. If \textit{\textbf{A}} type is ``home”, the agent residential location is used. Else, a location where \textit{\textbf{A}} can be performed has to be drawn, i.e. a shop or sustenance facility for a “shopping” activity, a healthcare facility for a “medical” activity, a residence for a “social” activity, etc. However, the TUS does not report activities exact locations. Moreover, the TUS is not trip oriented and thus does not contain trip distance between activities. The trip distance of \textit{\textbf{T-1}} from one activity \textit{\textbf{A-1}} to the next activity \textit{\textbf{A}} has to be estimated using the \textit{\textbf{T-1}} trip's mode and duration.
    
    An average speed has been assigned to each mode ([car: 30km/h, ride: 30km/h, public transit: 25/h, bike: 15km/h, walk: 4.8km/h]) according to Winnipeg observed speeds by mode \citep{winnipeg_tomtom} and to the observed general walking speed \citep{bosina_estimating_2017}.
    The walking, biking and driving speeds are assumed to be the same for all agents, even if this may be a bit too fast for the older people. With more data available than just average speeds, the older people could have been separated from the general population when making a location choice by assuming a reduced travel speed.
    
    To account for the 5 min rounding in TUS records and for the uncertainty in travel speeds, a trip distance range is calculated for \textit{\textbf{T-1}} rather than a fixed trip distance, according to:
        
    \begin{equation}
      \label{eq:t}
      \begin{aligned}
         min\_trip\_distance = speed(\textit{\textbf{T-1}}.mode) \times (\textit{\textbf{T-1}}.duration - 2.5)\\
        max\_trip\_distance = speed(\textit{\textbf{T-1}}.mode) \times (\textit{\textbf{T-1}}.duration + 2.5)
      \end{aligned}
    \end{equation}
    
    with distances in meters, speed in m/min and duration in minutes.

    Next, a location for \textit{\textbf{A}} is picked from the locations of the desired type that are within [$min\_trip\_distance$; $max\_trip\_distance$] from the previous activity \textit{\textbf{A-1}}. Because we assume all agents start their day at home, all activity locations can be incrementally estimated using this process.
    The reachable locations are identified using the agent previous location (\textit{\textbf{A-1}} location) and the computed distance range. However, individuals are traveling on the road network, and computing location-to-location road distances for 320730 residences and facilities is very computationally expensive. Previous research by- \cite{shadid_2009} found that Minkowski distance with a coefficient of 1.54 best approximates road distance. To identify potential locations for \textit{\textbf{A}}, we thus searched for locations of the desired type situated in a Minkowski distance (coefficient 1.54) between $min\_trip\_distance$ and $max\_trip\_distance$ from \textit{\textbf{A-1}}. One of the potential location is then randomly drawn and assigned for \textit{\textbf{A}}.

    \item The \textit{\textbf{A}} activity end time is randomized by +-15 min so that the 5 min rounding does not appear in the schedules and the agents’ schedules are more varied.

    \item The \textit{\textbf{A}} activity, defined by its end time, type (home, work, education, shopping, medical, eat out, social, entertain, exercise, escort someone) and location, and the \textit{\textbf{T}} trip, defined by its transportation mode (car, ride, public transport, walk, bike), are added to the agent schedule. When all trips have been added for the agent, its last activity (i.e. home) is added to terminate the schedule.
\end{enumerate}

% for indiv in syn_pop:
%         activities= indiv.get episodes:
%         trips = activities where type==”travel”

%         last_location = indiv.home
%         last_trip = first indiv trip

%         for trip in trips:
%             if consecutive trips identified, merge their durations
%             previous_activity = get_activity_type(trip-1, activities)
%             facility = get_facility(indiv, last_trip, previous_activity, last_location, facilities_by_type)
%             last_location = facility
%             last_trip = trip
%             end_activity = get_randomized_time(trip['STARTIME'])
% 	 mode = get_mode(trip['LOCATION'])

At the end of the modelling process, each synthetic individual with individual characteristics, is an agent with individual daily behaviour, in a digital Winnipeg city.

\section{Simulation and calibration with MATSim}\label{simulation}
The simulation phase of the study predicts the interactions and behaviours of the individual agents in the city throughout a day. In our case, we are interested in simulating the activity and mobility of the entire population, including the viability of services for older adult needs.
For the simulation, the Multi-Agent Transport Simulation (MATSim) framework \citep{horni2016introducing} was used (version 15.0). MATSim is a widely used open-source simulation framework designed to model and simulate large-scale transportation systems. By simulating the decision-making processes of thousands of agents, MATSim allows to investigate various transportation scenarios, and can help to assess the impacts of a new DRT service for older adults. 

The agents schedules have been converted to MATSim plans format. In addition to its plan, each agent has a $car\_availability$ attribute depending on its driving license ownership and a $subpopulation$ attribute to indicate if it is a older adult or not. At each time step, each agent executes its individual schedule, i.e. performs an activity at a specific location or travels from an activity to its next activity.
MATSim uses an evolutionary algorithm where each agent adapts its behaviour to the transport environment and maximise the utility of each plan. At each iteration, a proportion of the agents try a new plan by changing their transportation mode, route, or activity time. When an iteration ends plans are scored, i.e. each plan utility is calculated. At the end of the simulation, a traffic equilibrium is reached and it ensures that each agent uses its highest-scoring plan to produce a realistic demand.

\subsection{Simulation configuration}
The ABM was simulated for 500 iterations for the system to reach equilibrium, with 25\% of all agents due to MATSim high computational cost. Downscaling is a very common practice in MATSim, and \cite{ben-dor_population_2021} found that sampled populations equal to or greater than 25\% preserve most urban traffic statistics.
Three repetitions with different random 25\% samples have been executed. This corresponds to 192,244 agents with 31,607 older adults in the first version, 192,349 agents with 31,944 older adults in version 2 and 191,979 with 31,848 older adults in version 3.
The road network capacity has been scaled accordingly (flow capacity factor of 0.25 and storage capacity factor of 0.5 according to MATSim best practices). The public transit service on a Tuesday has been used and the access and egress time per person, number of seats, and bus’ passenger car equivalent have been reduced accordingly.

The simulation core used is \textit{Hermes}, a high-performance simulator for large-scale scenarios, that functions as a faster alternative mimicing most of the MATSim default simulator \textit{Qsim} functionalities \citep{hermes2021}.
The model is simulated second by second for 24h on a weekday (plus 6h to let late travelling agents finish their activities).
The car and ride users are routed on the network, and their travel times are impacted by congestion. However a ``ride” does not increase links loads as it is supposed to be shared with a ``car” trip.
Bike and walk trips are teleported, as these modes travel times are independent from the network congestion. For each bike or walk trip, a teleportation route is generated whose travel distance is 1.3 times the trip beeline distance (1.54 has been used for the location choices all modes considered, 1.3 is usually used for walk / bike as small shortcuts can apply for these modes), and whose travel time is the distance divided by the mode constant speed. The \textit{SwissRailRaptor} router is used for a fast public transit routing, and buses run with a steady speed according to their timetable (\textit{deterministicPT}).

At each iteration, 40\% of agents select their highest score plan from the previous iterations for re-execution in the current iteration (\textit{BestScore}), 20\% keep the same schedule and modes but try a different route (\textit{ReRoute}), 20\% shifts their activity end times randomly within a range of [-30min;+30min] (\textit{TimeAllocationMutator}), and 20\% change their mode of transport (\textit{SubtourModeChoice}). Agent can switch between car, ride, walk, bike and public transit, considering their driving license availability, i.e. an agent with no driving license can not use a car as a driver. Plan innovation is disabled for the last 10\% iterations, i.e. after 450 iterations the agents will only use plans from memory. This prevents agents from using unrealistic plans in the last iterations due to trying a new mode or departure time.
Score statistics and mode choice statistics have been checked and converge by the end of the iterations, showing that the equilibrium has been reached after 450 iterations.

All other parameters are the default ones: an A* landmark-based routing algorithm is used, agents can keep up to 5 plans in memory and the plan with the lowest score is deleted when the agent’s memory is full, etc.

Simulations were performed using the Linux-based High Performance Computing facilities from the University of Leeds. Each simulation has been executed on 24 cores (2.2GHz Broadwell E5-2650v4 CPUs with a memory bandwidth of 800MHz/core) and using 120GB RAM. Simulation computing time was 27h, i.e. 3 min 30 per iteration.

\subsection{Calibration process}
MATSim models require calibration by adjusting the mode utility parameters to ensure that the simulated modal split closely align with observed real-world data. Quantitative data on older adults' trips modal split in Winnipeg is limited, so the modal split for the global population was used for calibration.

The most recent travel survey for Winnipeg is the 2007 Winnipeg Area Travel Survey (WATS) \citep{wats_2007}, representing 4.4\% of all households within Winnipeg. All the daily trips made by the respondent householders (11 years and older) were recorded. Although the survey relates to travel from 15 years ago, it provides valuable insights into local travel behaviours. The 2007 WATS reports that among Winnipeg City households, car trips accounted for 64.8\% of the daily trips, ride for 15.8\%, public transit for 8.2\%, walk trips for 9.5\% and bike trips for 0.7\%.

The 2022 ABM modal split before calibration (just after the statistical matching) is close to the one reported by the 2007 WATS: 68.6\% of trips by car (+1.8\%), 14.1\% (-1.7\%) by ride, 9.5\% (+1.3\%) by public transit, 7.1\% (-2.4\%) by walk and 0.8\% (+0.1\%) by bike. To get an idea of how realistic are the differences in the modal split, we analysed the evolution of the main mode of commuting in Winnipeg reported between 2006 census \citep{census_2006} and 2021 census \citep{census_2021}. This confirmed that the car mode has been increasingly used (+3.5\% from 2006 to 2021), walk has been less used (-2\% from 2006 to 2021) and bike modal share has been constant (+0\% from 2006 to 2021). The 2022 ABM modal split before calibration is thus realistic for car, walk and bike modes.
However, while the ABM shows less ride and more public transit than in the 2007 WATS, the commuting ride modal share observed was constant (+0.3\% from 2006 to 2021), and the commuting public transit modal share observed decreased (-4.2\% from 2006 to 2021). The 2022 ABM modal split is thus a bit more uncertain regarding ride and public transit modes.
The modal split objective for the calibration was thus fixed as the input modal split with a $\pm$1\% tolerance for each mode ($\pm$1.2\% for ride and pt modes). 

\begin{figure}[ht]
    \centering
     \includegraphics[trim=0cm 0cm 0cm 0.9cm, clip,width=0.6\textwidth]{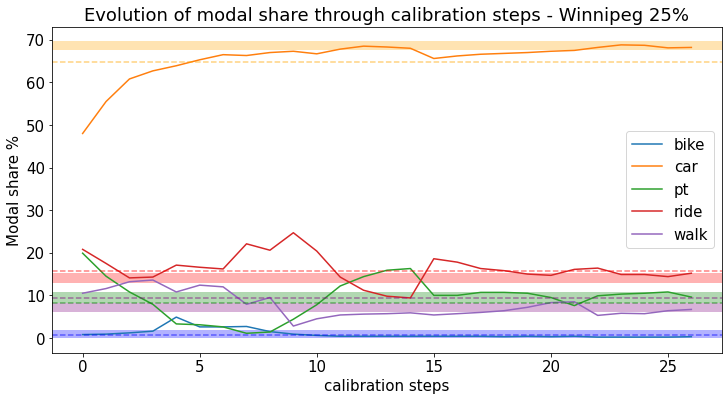}
     \caption{Iterative calibration of the simulated modal split for 25\% population sample (v1). The coloured ranges represent the modal share objective for each mode. The coloured lines represent the evolution of the modal share of each mode through the calibration steps. The dotted lines show the 2007 WATS modal share for each mode. After 27 steps, the modal share for each mode falls within the objective range.}
     \label{fig:fig5}
 \end{figure}

The modal split was calibrated using the first version of the 25\% sample model through an iterative process. At each simulation, the final simulated modal split (after 500 iterations) was analysed and the mode choice parameters were adjusted to push agents onto different modes, until a satisfactory match between the simulated and observed data was achieved. A first simulation was run with the default scoring parameters and teleportation speeds. The mode choice parameters have then been adjusted through simulations by altering the mode-specific constant, marginal utilities of distance and traveling in the scoring parameters, and the bike and walk teleportation speeds. 27 calibration steps were required to achieve a realistic modal split (Figure \ref{fig:fig5}).

The modal split of the calibrated model after 500 iterations, is presented in Table \ref{tab3}, along with the calibrated parameters for each mode. The public transit and ride modes have been assigned a large disutility in order to achieve realistic modal shares. To prevent very long walk or bike trips, larger negative utilities of travelling per meter and per hour have been set for these modes and their teleportation speeds have been reduced. All other scoring parameters (\textit{dailyMonetaryConstant}, \textit{dailyUtilityConstant} and \textit{monetaryDistanceRate}) have been set to zero.
It should be noted that the constants selected to obtain a realistic modal share of the ride, public transit, and bike are high, highlighting the economical or behavioral inconvenience of using these modes in Winnipeg, relative to the ease of driving a car in this context. The calibration of modal constants is a first step in designing a MATSim model. A more refined calibration of the utility of distance, traveling, and costs is then performed to obtain a complete decision model. Alternative approaches involving calibration of a choice model exogenous to MATSim are feasible, but outside the scope of this exploratory study.

\begin{table}[h]
\caption{Objective modal share, calibrated modal share in the simulated trips, and calibration parameters for each mode}\label{tab3}

\renewcommand{\arraystretch}{0.9} 
\begin{tabular*}{\textwidth}{@{\extracolsep\fill}lcccccc}
\toprule%
&& & \multicolumn{3}{@{}c@{}}{Scoring parameters} & \\
 \cmidrule{4-6}
Mode & Objective & Calibrated &constant &marg. utility&marg. utility&Teleportation\\
& (\%)&share (\%)&&of distance&of traveling&speed (m/s)\\
\midrule
car &[67.6, 69.6] &68.2& -10 & 0   & -6 & - \\
ride     & [12.9, 15.3]&15.2& -210 & 0   & -6 & - \\
public transit &[8.3, 10.7]&9.6& -225 & 0   & -6 & - \\
walk      &[6.1, 8.1]&6.7& 0  & -0.04 & -6 & 1 \\
bike     &[0, 1.8] &0.3& -50 & -0.04 & -36 & 2\\
\botrule
\end{tabular*}
\end{table}

\section{Validation}\label{validation}
Validating the ABM allows us to assess the reliability and accuracy of the model's predictions and outcomes, ensuring that the simulation adequately represents Winnipeg population and mobility. 

\subsection{Synthetic population and environment model validation}
\begin{figure}[ht]
    \centering
    \begin{subfigure}[b]{0.43\textwidth}
         \centering
         \includegraphics[trim=0cm 0.5cm 0cm 0.5cm ,clip,width=\textwidth]{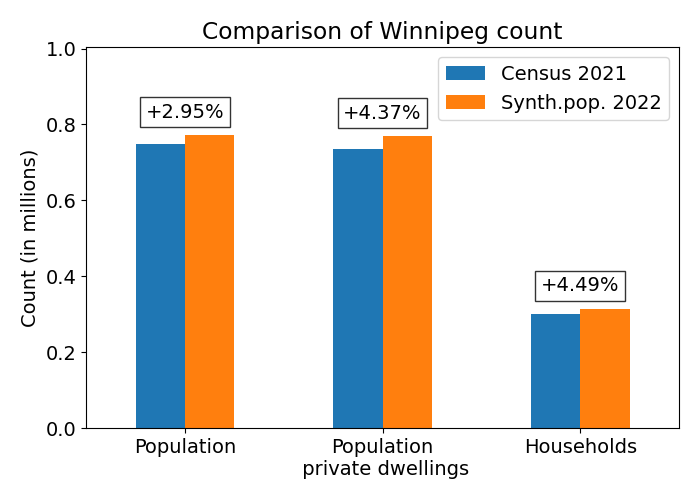}
         \caption{}
         \label{fig:fig6a}
     \end{subfigure}
     \begin{subfigure}[b]{0.39\textwidth}
         \centering
         \includegraphics[trim=0cm 0.5cm 0cm 0.5cm ,clip,width=\textwidth]{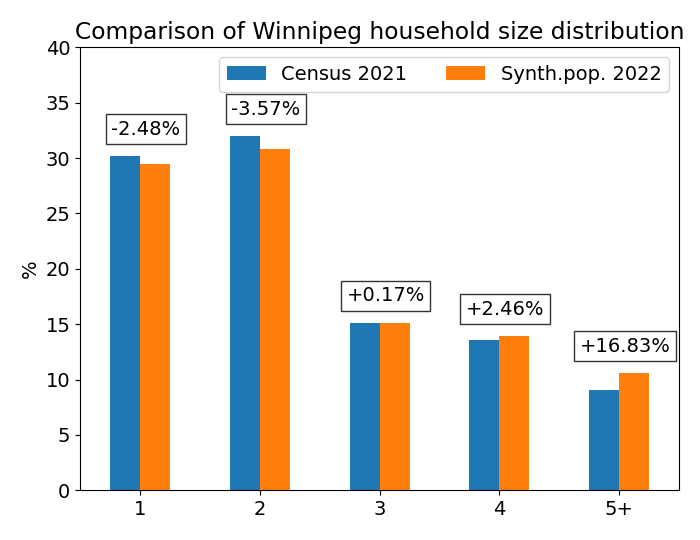}
         \caption{}
         \label{fig:fig6b}
     \end{subfigure}
     \begin{subfigure}[b]{0.98\textwidth}
         \centering
         \includegraphics[trim=0cm 0.5cm 0cm 0.5cm ,clip,width=\textwidth]{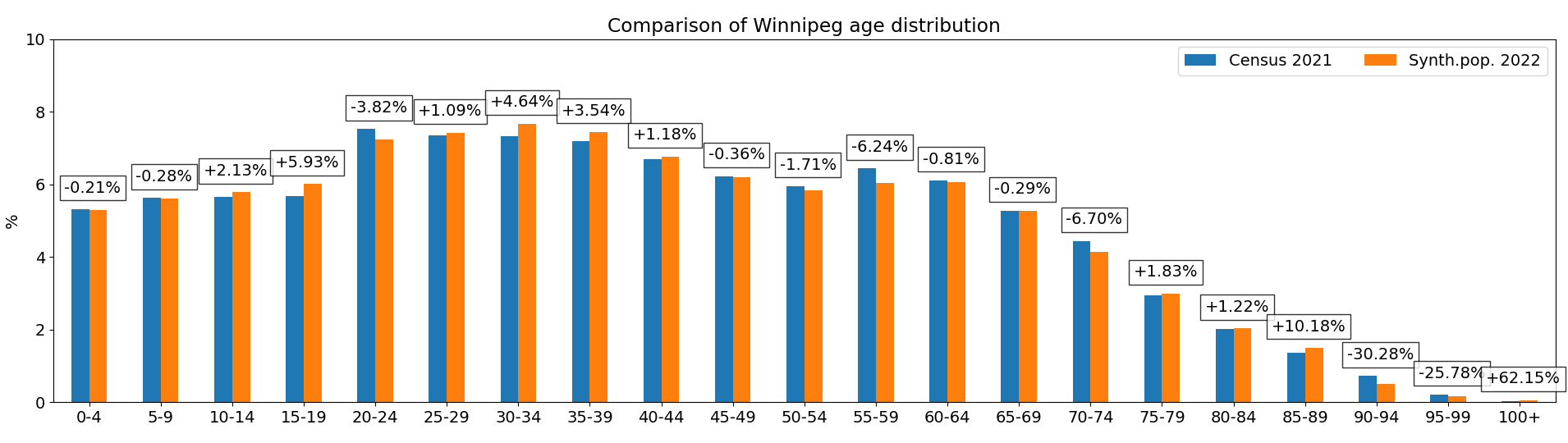}
         \caption{}
         \label{fig:fig6c}
     \end{subfigure}
    \caption{City level validation: comparison of 2022 synthetic population (in orange) and 2021 census population (in blue) for Winnipeg on (a) the population and households counts, (b) the household size distribution and (c) the age distribution. Relative errors appear in boxes for each category.}
    \label{fig:fig6}
\end{figure}

The synthetic population generation process has been previously validated for 2021 by \cite{predhumeau_2023} by comparing a synthetic population generated for 2021 to the actual 2021 population reported by the census. A city level validation has been presented for Winnipeg, with histograms and relative errors for each attribute and each category. 
A similar validation for the 2022 synthetic population is presented in Figure \ref{fig:fig6} regarding Winnipeg population distribution of sex, age and household size. 

Figure \ref{fig:fig6} shows that the 2022 synthetic population (in orange) accurately reproduces the 2021 census statistics (in blue) at the city level, with a larger population in the 2022 model compared to what was measured in 2021. The synthetic population for 2022 also includes the demographic shift due to the population ageing.

Additional validation has been presented in our previous work \citep{predhumeau_cupum_2023} for 2023, where the Winnipeg synthetic population density and mean age by DA have been compared to those reported by the 2021 census. The comparison showed that for 90\% of the DAs, the absolute difference in density is less than 20\% and the difference in the mean age is less than 4 years. The road network model completeness, and the residential and facilities counts have been validated as well.

\begin{figure}[ht]
    \centering
    \begin{subfigure}[b]{0.48\textwidth}
         \centering
         \includegraphics[width=\textwidth]{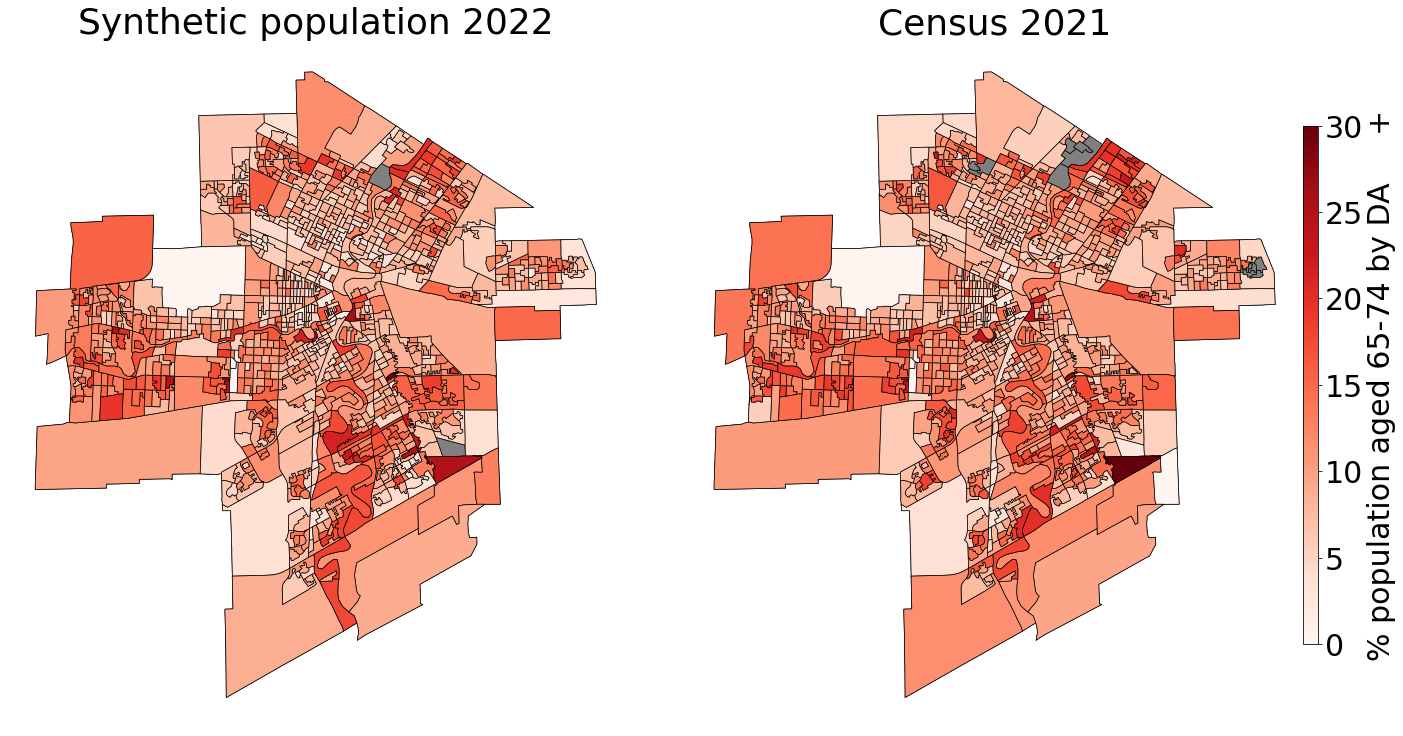}
         \caption{}
         \label{fig:fig7a}
     \end{subfigure}
     \begin{subfigure}[b]{0.48\textwidth}
         \centering
         \includegraphics[width=\textwidth]{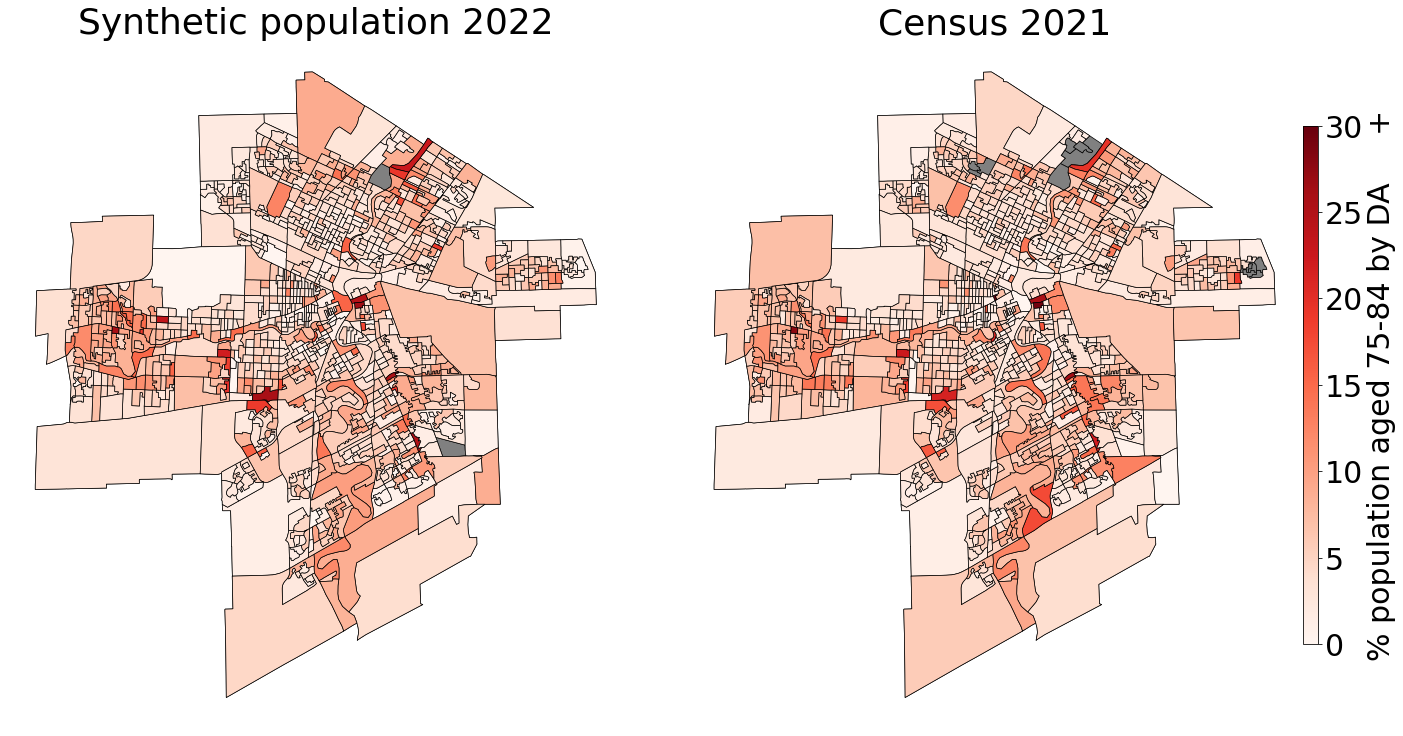}
         \caption{}
         \label{fig:fig7b}
     \end{subfigure}
     \begin{subfigure}[b]{0.48\textwidth}
         \centering
         \includegraphics[width=\textwidth]{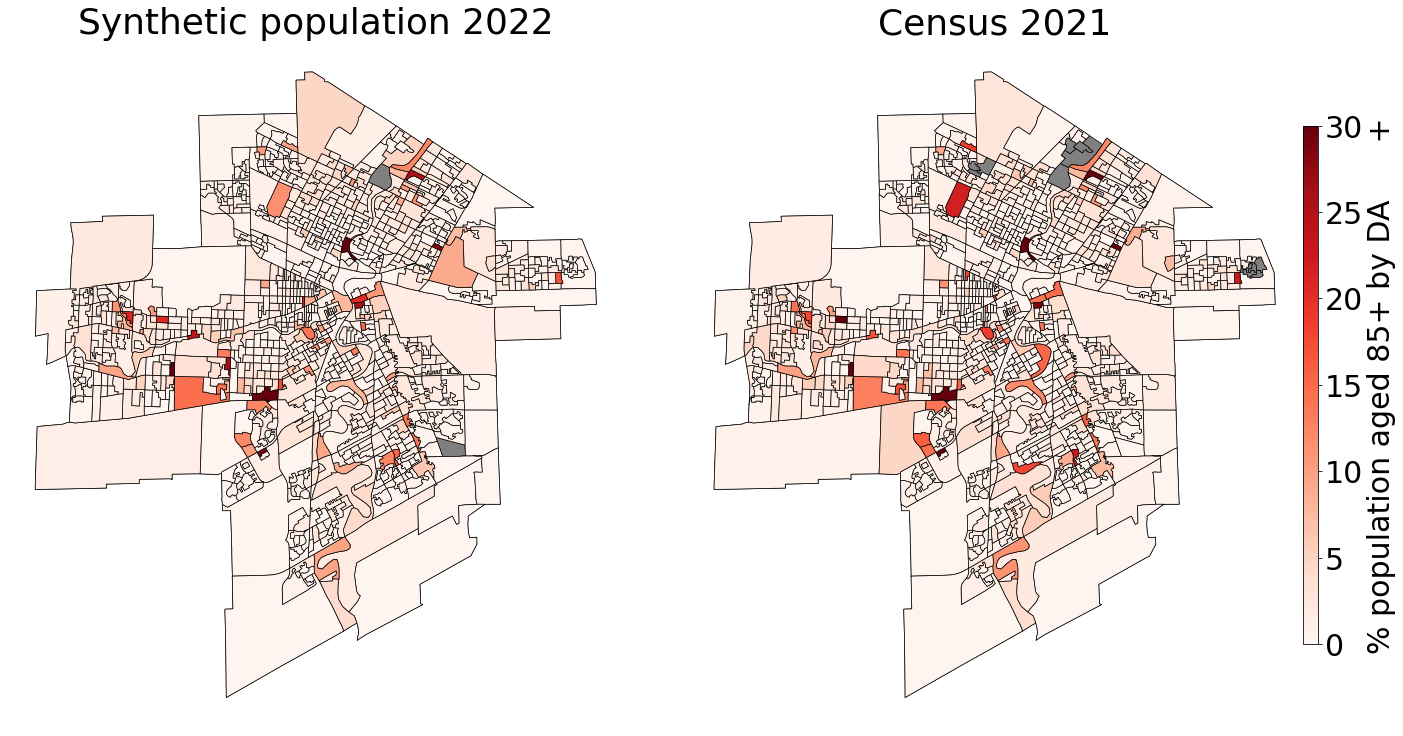}
         \caption{}
         \label{fig:fig7c}
     \end{subfigure}
    \caption{Dissemination area level validation for older adults: comparison of the 2022 synthetic older adult population (on the left-hand side) and the 2021 census older adult population (on the right-hand side) for Winnipeg. Proportions of people aged (a) 65 to 74, (b) 75 to 84 and (c) 85+ by DA. Grey DAs are not populated or not directly comparable due to boundaries update since last Census.}
    \label{fig:fig7}
\end{figure}

Finally, the synthetic older adult population spatial distribution was validated.
Figure \ref{fig:fig7} presents a comparison of the 2022 synthetic older adult population with the 2021 census older adult population at the DA level. The proportions of people aged 65 to 74, 75 to 84 and 85+ by DA from the census are well reproduced by the synthetic population.
The mean absolute proportion error over all DAs is 2.1\% for the 65 to 74 year-old (95\% of DAs have an error $<$6\%), 1.6\% for the 75 to 84 year-old (95\% of DAs have an error $<$5\%), and 1\% for the 85+ year-old (95\% of DAs have an error $<$3\%). As they age, the elderly populations become more prevalent in specific neighbourhoods: in St. James-Assiniboia West, in Assiniboia South, in the River East West neighbourhood (north) and in St. Vital along the Red river and Seine river.

\subsection{Activity-based model and mobility validation}
Various data sources have been used to compare the ABM outputs against real-world observations. Due to limited data regarding older adults' trips in Winnipeg, the adopted approach involved quantitatively validating the global population mobility before verifying that the mobility trends of older adults compared to the general population were accurately replicated. All the following figures report the mean values measured on the 3 simulation repetitions and the minimum and maximum values as black error bars.

\subsubsection{Global population mobility validation}
\begin{figure}[ht]
    \centering
    \begin{subfigure}[b]{0.48\textwidth}
         \centering
         \includegraphics[width=\textwidth]{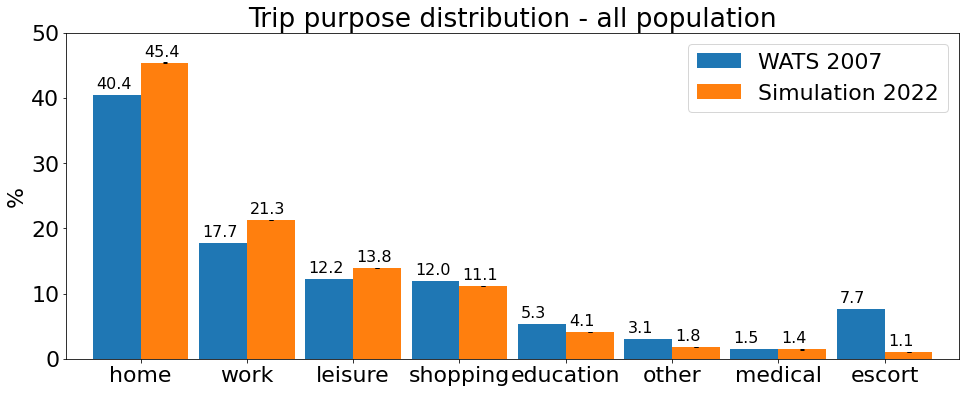}
         \caption{}
         \label{fig:fig8a}
     \end{subfigure}
     \begin{subfigure}[b]{0.48\textwidth}
         \centering
         \includegraphics[width=\textwidth]{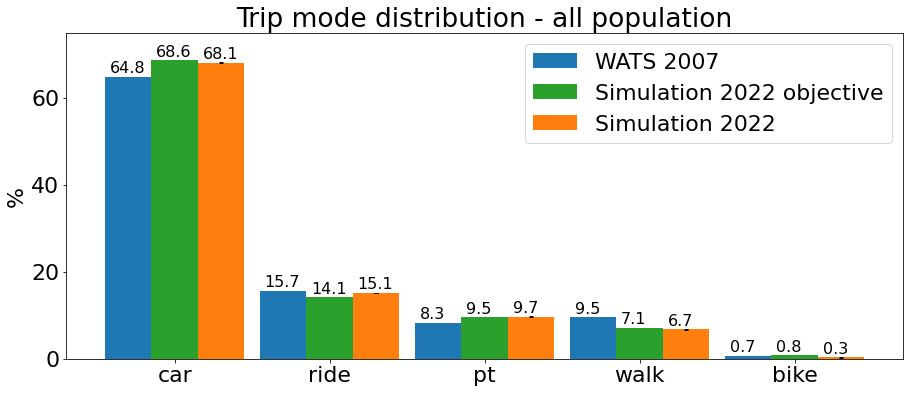}
         \caption{}
         \label{fig:fig8b}
     \end{subfigure}
     \begin{subfigure}[b]{0.48\textwidth}
         \centering
         \includegraphics[trim=0cm 0cm 0cm 0.2cm ,clip,width=\textwidth]{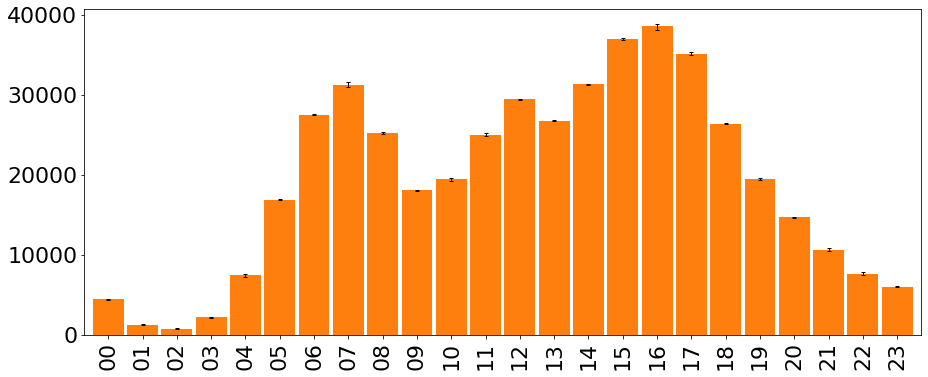}
         \caption{}
         \label{fig:fig8c}
     \end{subfigure}
     \begin{subfigure}[b]{0.48\textwidth}
         \centering
         \includegraphics[trim=14cm 2cm 12cm 18cm ,clip, width=\textwidth]{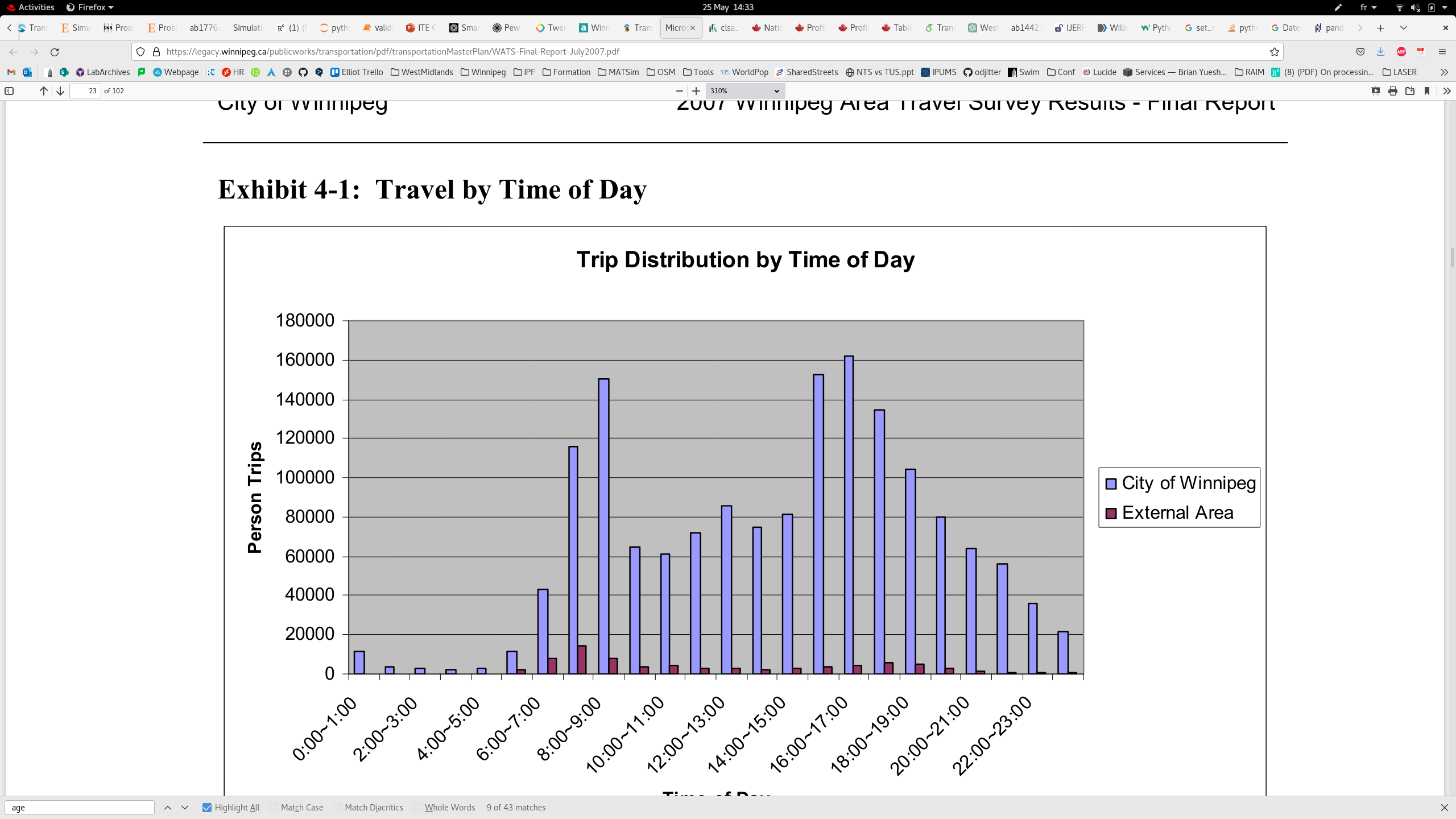}
         \caption{}
         \label{fig:fig8d}
     \end{subfigure}
    \caption{Distributions of global population (average over the three 25\% population samples) simulated trips' (a) purpose, (b) transportation mode, and (c) start time, in orange, compared to 2007 WATS distributions in blue. (d) Distribution of start times of trips by time of day, for the City of Winnipeg and the external area separately, from 2007 WATS \citep{wats_2007}.}
    \label{fig:fig8}
\end{figure}

Distributions of simulated trips' purposes, transportation modes and start times for the global population are quantitatively compared against distributions reported by the WATS 2007 \citep{wats_2007} in Figure \ref{fig:fig8}.

The trips purposes distribution in simulation reproduces well the 2007 WATS distribution; the main trip purposes are going back home, work, leisure (i.e. socialising, eating out, exercising and entertaining), and shopping (Figure \ref{fig:fig8a}). More escort trips are reported by the WATS 2007 than in the simulation. However, the low rate of escort trips in the model is coherent with the TUS data in which only 3.3\% of the respondents reported time spend accompanying a child, 0.7\% accompanying a teenager and 0.8\% accompanying an adult. The difference might be linked to differences in the way the WATS and TUS surveys have been conducted.
 
The simulated modal split is very close to the WATS modal split and to the objective defined in Section \ref{simulation}, as shown in Figure \ref{fig:fig8b}. This is expected as the simulated modal split was adjusted during the calibration step. The model was calibrated using the first version of the 25\% population and the very small error bars in Figure \ref{fig:fig8b} show that the two other versions are well calibrated using the same parameters. In Winnipeg, the dominant transportation mode is car as a driver, followed by car as a passenger, public transit, walk and then bike, which the model accurately replicates.

The simulated trips distribution by time of the day (Figure \ref{fig:fig8c}) have been compared to the WATS distribution (Figure \ref{fig:fig8d}). Once again the model reproduces very well the morning and afternoon traffic peaks observed in Winnipeg with the highest traffic peak from 4 p.m. to 5 p.m.

Aggregated trips statistics are well reproduced by the model. 
While the 2007 WATS reports 2.83 daily trips per person on average (for people aged 11 and over), the proposed model has an average of 2.69 daily trips per agent aged 11 and over.
The mean commuting time reported by the 2016 census \citep{census_2016} for Winnipeg is 24 min while the ``work'' trips in the model have an average duration of 22 min. We found no available data on the trip distances in Winnipeg. However, a quarterly report on national vehicle kilometres travelled metrics for 2021 by Automotive Industries Association of Canada and StreetLight Data \citep{aia_2021} reported a national average trip length of 15 km. The average trip length by car (or ride) in the simulation is close; 13 km.

\subsubsection{Older adults mobility validation}
After validating the global population simulated mobility, older adults mobility trends has been validated comparatively and using national studies.
As reported by \cite{li_population_2012}, older adults travel less as they age. This is well reproduced by the model; agents aged 11 to 64 perform 2.72 daily trips on average, agents aged 65 to 74 perform 2.66 daily trips on average and agents aged 75 and over perform 2.42 daily trips on average.

\begin{figure}[ht]
    \centering
    \begin{subfigure}[b]{0.48\textwidth}
         \centering
         \includegraphics[width=\textwidth]{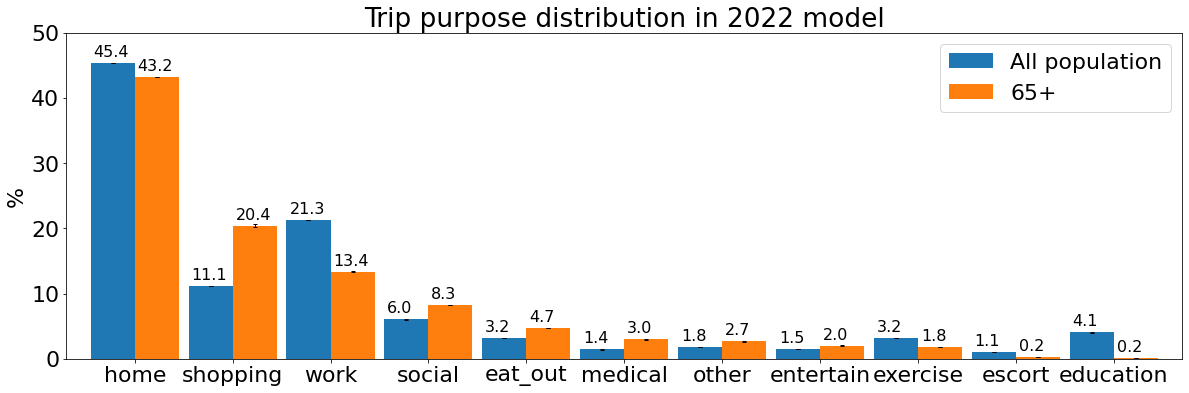}
         \caption{}
         \label{fig:fig9a}
     \end{subfigure}
     \begin{subfigure}[b]{0.48\textwidth}
         \centering
         \includegraphics[width=\textwidth]{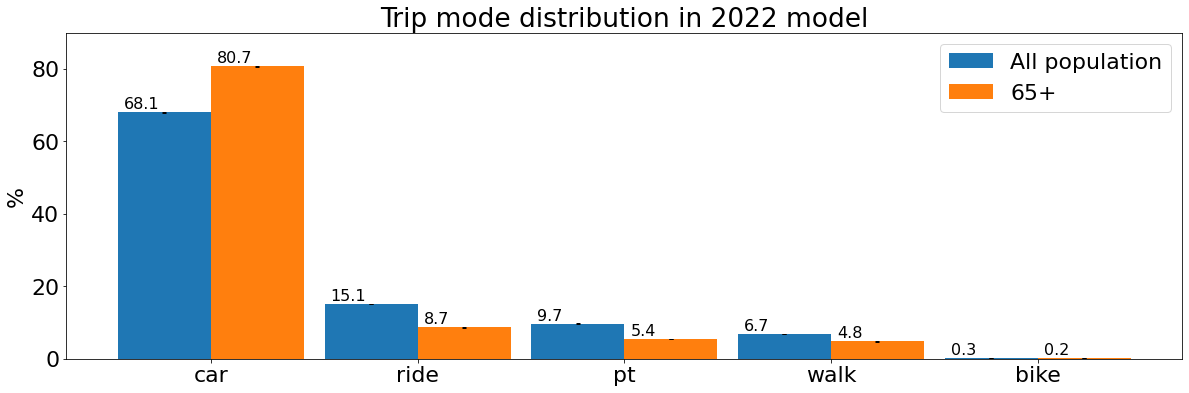}
         \caption{}
         \label{fig:fig9b}
     \end{subfigure}
     \begin{subfigure}[b]{0.48\textwidth}
         \centering
         \includegraphics[trim=0cm 0cm 0cm 0.2cm ,clip,width=\textwidth]{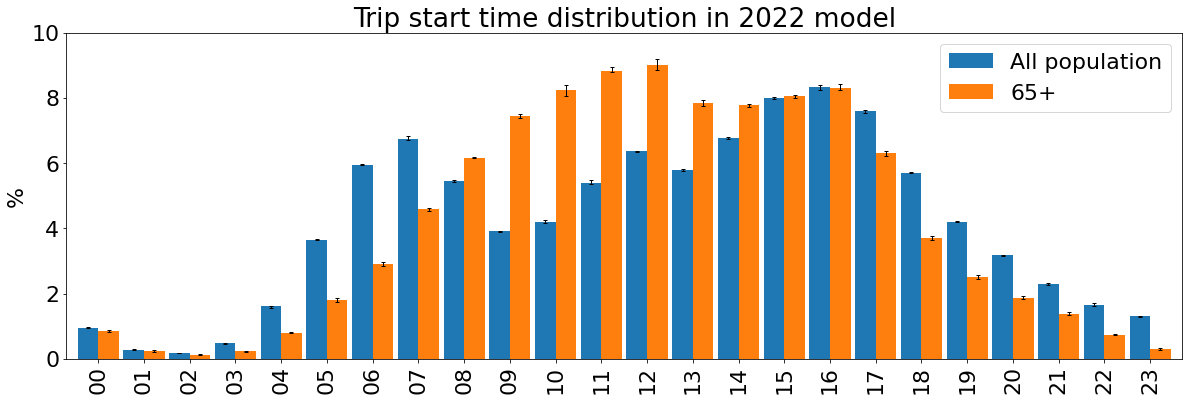}
         \caption{}
         \label{fig:fig9c}
     \end{subfigure}
    \caption{Distributions of simulated trips' (a) purpose, (b) transportation mode, and (c) start time, for the global population (in blue) and seniors population (in orange). Average values over three simulation repetitions are presented, with min and max values as error bars.}
    \label{fig:fig9}
\end{figure}

Compared to the general agent population, simulated older adults have more ``shopping'', ``social'', ``eat out'', and ``medical'' trips and less trips related to ``work'', ``exercise'' or ``education'' (Figure \ref{fig:fig9a}). These trends have been observed in real older adult populations by several studies from different countries \citep{fatima_elderly_2020,li_population_2012, watson_2021, mohammadian_2013, ohern_understanding_2015}. After 65, an increasing number of people retire and thus have less commuting trips, education trips are mainly performed by children and young adult students, and with age physical abilities may decline and hamper exercise. Still, in 2015, 19.8\% Canadians aged 65 and older, worked at some point during the year \citep{canada_senior_worker}, which explains the high percentage of trips related to work in this population.

\begin{figure}[ht]
    \centering
    \begin{subfigure}[b]{0.48\textwidth}
         \centering
         \includegraphics[width=\textwidth]{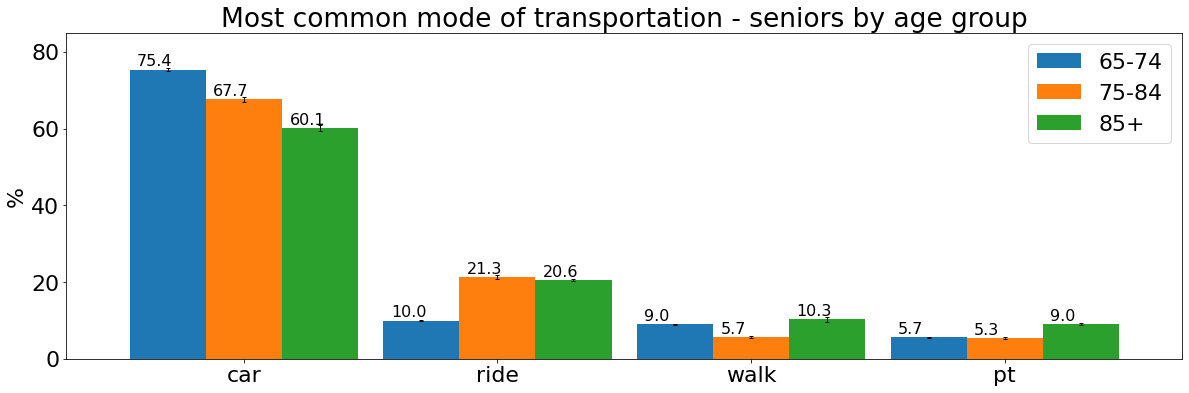}
         \caption{}
         \label{fig:fig10a}
     \end{subfigure}
     \begin{subfigure}[b]{0.48\textwidth}
         \centering
         \includegraphics[width=\textwidth]{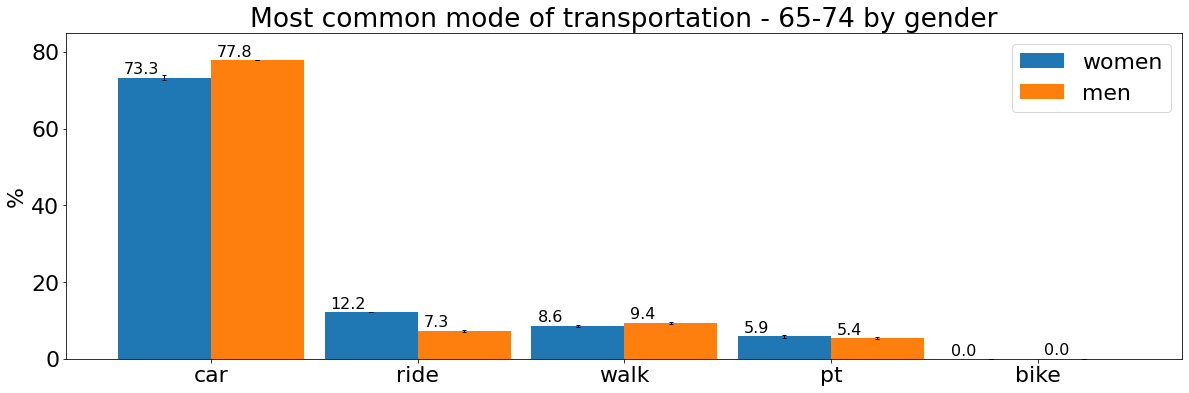}
         \caption{}
         \label{fig:fig10b}
     \end{subfigure}
    \caption{Distributions of simulated older adults' trips mode (a) by age range, and (b) by gender. Average values over three simulation repetitions are presented, with min and max values as error bars.}
    \label{fig:fig10}
\end{figure}

Older adult mobility is heavily dependent on car use \citep{luiu_travel_2021, li_population_2012, ohern_understanding_2015}. This is well reproduced by the model; 80.7\% of older adults' trips are by car (Figure \ref{fig:fig9b}).
However, observations show that the modal split is not homogeneous among older adults \citep{li_population_2012, ofallon_2009}. In Canada, national statistics on elderly mobility are reported by the Canadian Longitudinal Study on Aging (CLSA) \citep{clsa}, a study of adult ageing on 50,000 Canadians aged 45 to 85 years, and by the Canadian Community Health Survey (CCHS) \citep{cchs_2009}, an annual cross-sectional survey that collects information related to health status, health system and health determinants for the Canadian population. These studies report two key observations on older adults travel: 1) as they age, older people use the car less as a driver and more as a passenger, and 2) older men use the car more as a driver and less as a passenger compared to older women.
These two observations are well reproduced in the model, as shown in Figure \ref{fig:fig10}.
With age, agents use less car as drivers and more as passengers (Figure \ref{fig:fig10a}). Among simulated older adults aged 65 to 74, men (in orange) drive more than women (in blue) and are less passengers (Figure \ref{fig:fig10b}).

Older adults have temporal mobility patterns different from the global population; according to \cite{ofallon_2009} most elderly people make their trips between 9:30 a.m. and 3 p.m. This also appears in the model where older adults' trips do not show a morning and an afternoon peak as the global population do, but rather a smooth increase in the morning until midday and a smooth decrease after 4 p.m (Figure \ref{fig:fig9c}).

Finally, an analysis of the TUS data shows that 29.9\% of older respondents stayed at home during their reported week day (15.9\% of general population respondents stayed at home). In the model, 20.8\% of older adults agents stay at home (i.e. no trip and activity other than ``home'') for the simulated day compared to 13.5\% considering all simulated agents.
This trend is very pronounced for older adults aged 85 and over; 46\% of 85+ agents stay at home for the simulated day. Among the 85 year-old and over, women agents stay more at home than men agents: 55\% vs. 30\%. Finally, the most significantly correlated attribute for staying at home is not having a driving license: 75\% of non-licensed 85+ agents stay at home vs. 8\% for the licensed 85+ agents. Similar findings have been previously observed for older New Zealanders \citep{ofallon_2009}.

% OD 2018 data comparison: we overestimate the nb of intra winnipeg trips by 30%. Can be explained by OD data are from 2018 and we simulate 2022 + during the trips generation, we assume that all trips are made within Winnipeg+smartphones. Compare for seniors?
% Main congestion points validated ?

\section{Preliminary testing of an AMoD service in simulation} \label{exploration}
% Results should be clear and concise.
Once the model was successfully validated, the focus shifted towards conducting preliminary testing of an AMoD service in the simulated environment.
% The preliminary testing aimed to serve as a proof-of-concept before a more complete exploration of various AMoD designs in future work.

\subsection{Simulation scenarios}
Several AMoD scenarios have been explored in order to be compared and to draw first insights into the potential older adults' demand:
\begin{itemize}
    \item The \textit{baseline scenario} is the calibrated scenario after 500 iterations representing the current Winnipeg mobility, with no AMoD service.
    
    \item The \textit{AMoD scenarios} are scenarios where the demand is the calibrated demand from the baseline scenario, and an AMoD mode has been added for older adults. Older adult agents can thus switch from their calibrated mode to AMoD and if this new mode is more convenient, they will use it.
    
    \item Finally, \cite{harb_2018} showed that the introduction of a new transportation service may generate some induced demand by new user groups. Induced demand refers to the phenomenon where the availability of a service leads to an increase in demand for that service. People who have difficulties using a car or public transport could now move more easily, and go out of home more often, using the door-to-door AMoD service, which thus might generate new trips.
    This new demand from underserved populations has been evaluated in previous works with various assumptions. \cite{harper_2016} assumed that with AVs the non-drivers would travel as much as drivers from the same age or younger. \cite{brown_2014} assumed that the population segments aged 16 to 85 would travel as much as the top decile of the segment. \cite{wadud_2016} estimated that AVs could lead to a share of drivers similar to the 35-55 age group across all age groups and that everyone aged 62 and above would drive as much as those 62 years old. Finally, \cite{hogeveen_2021} assumed that longer and more frequent trips would be made with AVs, and that new user groups like children or the elderly would use AVs. 

    Inspired by these previous approaches, the \textit{AMoD induced scenarios} explore situations where the demand is the calibrated demand from the baseline scenario, but modified so that older adults with no car access have now the same mobility patterns as older adults with car access.
    A new approach has been used; the non-licensed older adults have been matched again with the TUS records, assuming that they have a license during the matching process. Their new activity plans have then been generated according to the procedure presented in Section \ref{acbm} and the ``car” mode (inaccessible because they do not have a driving license) has been replaced by the AMoD mode in their plans. The new demand, including the induced demand for older adults, is used for simulations where an AMoD mode has been added to the available modes for older adults. All older adult agents can thus switch from their calibrated mode to AMoD.
\end{itemize}

Moreover, the fleet size is an important element for the AMoD adoption, as a too small fleet will not be able to serve all the requests. We thus simulated each AMoD scenario with 3 different fleet sizes: 100, 250 and 500 vehicles.
Each scenario with AMoD has been run with the default MATSim core (QSim) for 150 iterations, and has been repeated for the three 25\% population sample versions. 150 iterations proved sufficient to reach the equilibrium point, with the score and mode statistics converging after 100 iterations. All innovations have been switched off for the last 10\% iterations.

All agents can try new modes but MATSim code has been modified so that only the older adult subpopulation has access to the AMoD mode.
Agents activity and mode scoring are kept as in the baseline scenario, and the AMoD mode is scored by agents similarly to the car mode.

\subsection{Simulated AMoD service}
The AMoD service has been simulated using MATSim DRT module \citep{Bischoff2017}, which allows the simulation of shared rides. For each incoming user request, it assess all possible insertion points in AMoD vehicles trips and chose the best one or reject the request. The AMoD vehicles operate in a door2door mode, and are routed on the road network. The fleet size is composed of 100, 250 or 500 vehicles depending on the scenario.
Each vehicle has a capacity of 4 passengers.
All vehicles are at the Winnipeg Transit bus depot (central location in Winnipeg city) at the start of the simulation. During the simulation, the default Minimum Cost Flow Rebalancing strategy is used to efficiently rebalance the vehicles and to take into account the demand structure of the previous iteration. Idle vehicles return to the nearest of all start links. 

Older users will wait for a maximum of 30 min after they submit their request (emulating an online or app request system) and each pick up or drop off stop will be 1 min. The waiting times is considered as a hard constraint; the request gets rejected if it can not be served on time, and the agent will not perform the trip and attribute a very low score to this plan.

All other parameters have been kept to the default values; maxTravelTimeAlpha: 1.5, travelTimeEstimationAlpha: 0.05, travelTimeEstimationBeta: 0, targetAlpha: 0.5, targetBeta: 0.5, zonesGeneration: GridFromNetwork, and maxTravelTimeBeta: maxWaitTime.
		
The future AMoD service would be provided with autonomous electric vehicles (EV), so the AV and EV MATSim modules have been used \citep{Bischoff2019}. The AV module allows to adapt the road capacity in order to reflect the AVs increased efficiency; two AVs would need only the flow capacity of one ordinary vehicle.
The EV module allows to simulate the electric vehicles battery, add charging stations and make the vehicles go to stations when their battery is low. The EV configuration has been kept to the default configuration; vehicles are sent to chargers when their battery is below 20\% and they charge up to 80\%. Winnipeg charging locations have been retrieved from the Alternative Fueling Station Locator provided by the Alternative Fuels Data Center \citep{afdc_2023}. It provides the location of the public electric charging stations in the United States and Canada.

Simulations have been performed on the Linux-based High Performance Computing facilities from the University of Leeds., with 24 cores (2.2GHz Broadwell E5-2650v4 CPUs with a memory bandwidth of 800MHz/core) and 120GB RAM.
Simulation computing time was 32h to 40h for the 100 vehicles fleet, 46h to 53h for the 250 vehicles fleet and 51h to 63h for the 500 vehicles fleet.

\subsection{Results and discussion}
\begin{figure}[ht]
    \centering
         \includegraphics[width=0.5\textwidth]{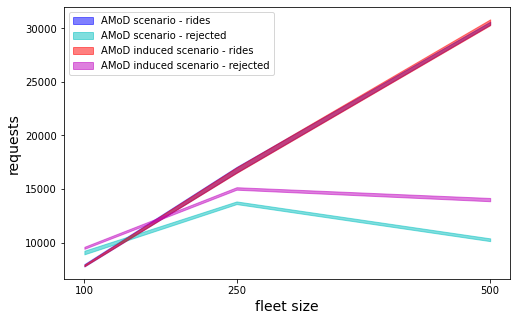}  
    \caption{Number of requests served (dark blue and red) and rejected (light blue and pink) in the AMoD and AMoD induced scenarios, with 100, 250 and 500 vehicles.}
    \label{fig:fig11}
\end{figure}

% How many requests and reject for each scenario: chart (impact of fleet size and induced demand)
We examined the relationship between the fleet size and the number of rides and rejections in the AMoD system, while also considering the presence or absence of induced demand (Figure \ref{fig:fig11}).
As the fleet size increased, we observed an increase in AMoD rides due to improved capacity and availability, suggesting that the demand adapts to the fleet size. 

The rejection rate is high with 100 AMoD vehicles; more requests are rejected than accepted (53\% to 55\% of requests are rejected), and thus many older adult agents do not switch to AMoD in their final plans.
With 250 AMoD vehicles, the number of requests increases and more rides are served; 45\% to 48\% of requests are rejected. More agents had a successful ride and thus switch to AMoD but the rejection rate is still high in the final iteration.
Finally, with 500 AMoD vehicles more rides are requested because many agents switched to AMoD and the rejection rate is lower (only 25\% to 32\% of requests are rejected).
While the number of rides served increases linearly with the fleet size, the number of rejected requests increases with 250 vehicles and decreases with 500 vehicles. This suggests that 500 vehicles is a minimum fleet size to start to absorb the demand but is not sufficient to serve all requests. This high rejection rate is related to long waiting times: 25\% of all requests are rejected after a 30 min wait. It appears that by increasing this waiting time limit, more requests could be satisfied.

When considering the induced demand, more older adults trips are happening: 2.82 daily trips on average for agents aged 65 to 74 instead of 2.66, and 3.17 daily trips on average for agents aged 75 and over instead of 2.42.
After the induced trips are considered, the number of daily trips for people above 75 years old increases more than that of the people between the age of 65 and 74. This result is due to the strong dependence between the number of daily trips and access to the car. While 65-74 year-old seniors without car access tend to maintain activities out of home using other transportation modes, seniors above 75 years old mobility is more impacted by the absence of car access. If an AMoD system can be used as a car, then older seniors might maintain a high level of mobility.

We noticed a similar number of AMoD rides served with the induced demand compared to without the induced demand. This is expected as the service is full and additional requests have to be rejected. This is coherent with the higher number of rejected requests when the induced demand is considered; the demand is higher with the induced demand but the fleet size is insufficient to support it.

\begin{figure}[ht!]
    \centering 
    \begin{subfigure}[b]{0.49\textwidth}
         \centering
         \includegraphics[trim=0.2cm 0.2cm 0.2cm 1.25cm ,clip,width=\textwidth]{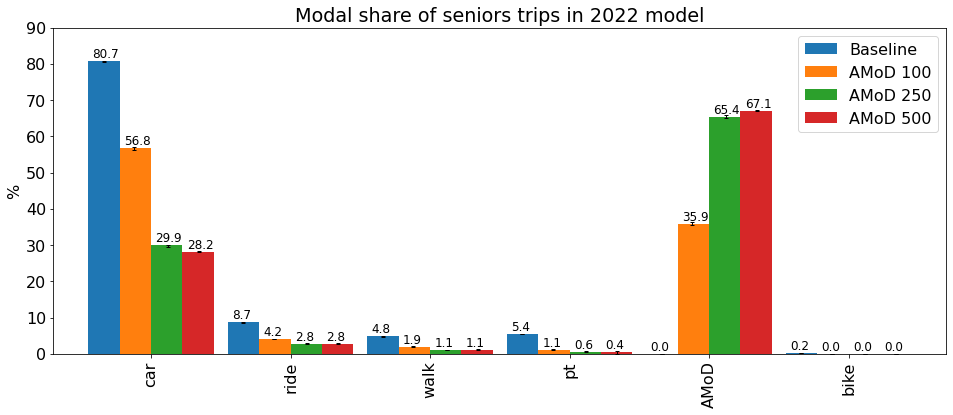}
         \caption{}
         \label{fig:fig12a}
     \end{subfigure}
     \begin{subfigure}[b]{0.49\textwidth}
         \centering
         \includegraphics[trim=0.2cm 0.2cm 0.2cm 1.25cm ,clip,width=\textwidth]{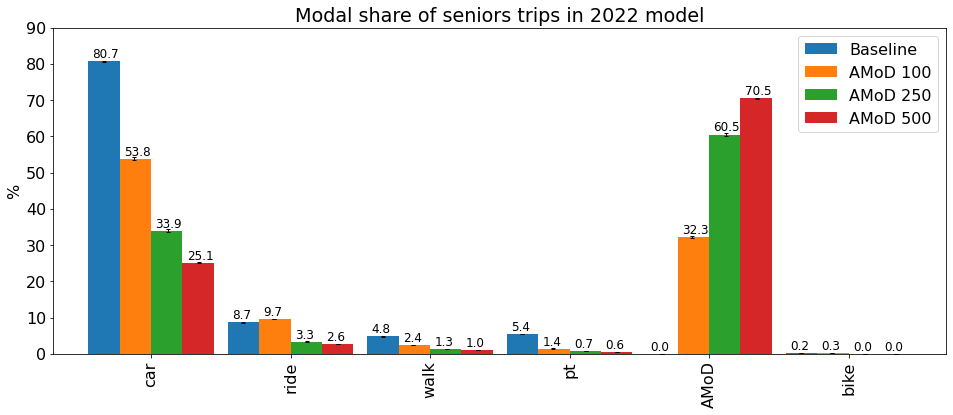}
         \caption{}
         \label{fig:fig12b}
     \end{subfigure}
    \caption{Older adults trips’ modal share for each scenario: baseline scenarios (blue), AMoD with 100 vehicles (orange), AMoD with 250 vehicles (green) and AMoD with 500 vehicles (red) (a) without the induced demand and (b) with the induced demand. Average values over the three repetitions are presented, with min and max values as error bars.}
    \label{fig:fig12}
\end{figure}

Regarding the modal share, older adults switched from all modes to AMoD (Figure \ref{fig:fig12a}).

With 100 vehicles, AMoD represents 35.9\% of the trips, while with 250 vehicles AMoD accounts for 65.4\% of the trips, becoming the main transportation mode for older adults.
Surprisingly, the modal split changes very little from 250 vehicles to 500 vehicles. This might indicate that almost all older adults who could switch to AMoD have already done so with 250 vehicles and that adding more vehicles will absorb the demand but not generate more demand.

The use of biking, public transit (pt), and walking as transportation modes has dwindled to marginal shares, with each of these modes representing less than 2\% of the trips. The use of car as a driver or a passenger is reduced from 89.4\% to 31\% of trips with 500 AMoD vehicles. Even if the AMoD mode utility is valued as the car mode, some agents stick to the car mode, due to the additional waiting time and detour time of the AMoD mode. 

Similar observations are made when considering the induced demand (Figure \ref{fig:fig12b}). However, the AMoD modal share continues to increase when the fleet size is increased from 250 vehicles to 500 vehicles. This means that the demand for AMoD is higher and 250 vehicles is not sufficient to absorb the potential demand. If more vehicles are added to the fleet, then more agents will switch to AMoD. This is coherent with the higher rejection rate observed on the 500 vehicles when the induced demand is considered: 32\% compared to 25\% without the induced demand.
Larger fleets have to be explored in order to identify the typing point where no more agents adopt the AMoD mode.

The AMoD users vary depending on the fleet size and on whether the induced demand is considered or not, as shown in Table \ref{tab4}.

\begin{table}[h]
\caption{AMoD users characteristics compared to the general older adult population, for each scenario. Proportions higher than the general population are in bold}\label{tab4}
\begin{tabular*}{\textwidth}{@{\extracolsep\fill}lcccccccc}
\toprule%
& Older adult  & \multicolumn{6}{c}{AMoD users} \\
&agents population&\\
\midrule
Induced demand & & \multicolumn{3}{c}{No} & \multicolumn{3}{c}{Yes} \\
 \cmidrule{3-8}
Fleet size & & 100  & 250  & 500  & 100  & 250  & 500  \\
\% women & 55.9  & 53.5  & \textbf{56.6} & \textbf{56.3} & 55.5 & \textbf{58} & \textbf{59.5} \\
\hline
\% license owners & 77.6 & \textbf{80.5}  & \textbf{85.5} &\textbf{ 85.9} & 62 & 66 & 69  \\
\hline
Mean age & 74.7 & 73.7  & 74  & 74  & \textbf{75.2} & \textbf{75.5} & \textbf{75.4} \\

\botrule
\end{tabular*}
\end{table}

Without considering the induced demand, only 53.5\% of the first users are women while they represent 55.9\% of the general older adults population. As the fleet size increases, the users gradually shift towards more women; with 500 vehicles 56.3\% of the AMoD trips are made by women.
When the induced demand is considered, the first users gender distribution is similar to the general older adults population gender distribution, meaning that more induced demand could come from women than from men. Again, as the fleet size increases, the users gradually shift towards even more women; with 100 vehicles, 55.5\% of AMoD trips are made by women, and with 500 vehicles 59\% of the AMoD trips are made by women.
Fewer women than men have access to a car, particularly in the older age groups, where many women do not go out from home. Many women who do not have access to a car may therefore use the AMoD service to make trips they cannot make without a car.

Without considering the induced demand, the primary users are the one with a driving license; while 77.6\% of older adults are licensed, 80.5\% of AMoD trips are made by licensed older adults with 100 vehicles. When the fleet size increases, more and more users have a driving license; with 500 vehicles 86\% of AMoD trips are made by licensed older adults.
When the induced demand is considered, more primary users do not have a driving license; with 100 vehicles only 62\% of AMoD trips are made by licensed older adults. Again, when the fleet size increases, more and more of the users have a driving license. However, even with 500 vehicles only 69\% of AMoD trips are by license owners. This is expected because the induced demand in the model comes from agents without driving license, which thus represent an important part of the AMoD users.

These findings indicate that while the first users might be men and non-licensed older adults, with a larger fleet, women and licensed older adults might switch to AMoD as well.

\begin{figure}[ht]
    \centering 
         \includegraphics[width=0.7\textwidth]{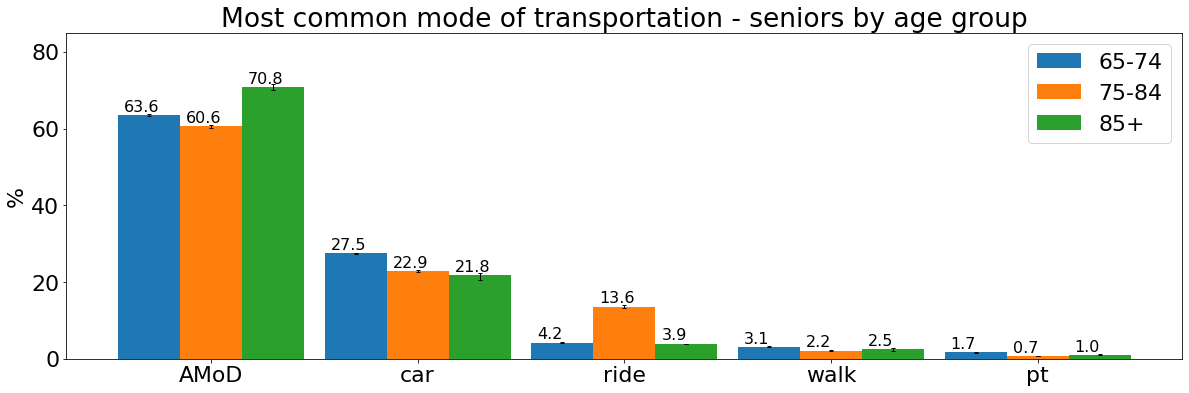}
    \caption{Older adults most common mode of transportation by age group, for the AMoD 500 vehicle scenario (no induced demand). Average values over the three simulation repetitions are presented, with minimum and maximum values as error bars.}
    \label{fig:fig15}
\end{figure}

Finally, the AMoD users in the non-induced demand scenario are a bit younger than the general population; their mean age is 74 years while the global older adults population mean age is 74.7 years. On the contrary, when the induced demand is considered, the users are a bit older than the general population; AMoD users mean age is 75.2 years old with 100 vehicles and 75.5 with 500 vehicles. The induced demand comes from non-licensed older adults, who are more prevalent in the oldest populations, and use AMoD for their trips.
Even when the induced demand is ignored, with a fleet of 500 vehicles, AMoD becomes the main mode of transportation for all older adult age ranges (Figure \ref{fig:fig15}). While it is the most common mode of transportation for 63.6\% of older adults aged 65 to 74, and for 60.6\% of older adults aged 75 to 84, it is the main mode of transportation for up to 70.8\% of the oldest segment (85+). 
Among these older adults, 64\% of car users, 76\% of walkers, 81\% of car passengers and 89\% of public transit users switched to AMoD.

\begin{figure}[ht]
    \centering
    \begin{subfigure}[b]{0.48\textwidth}
         \centering
         \includegraphics[trim=0cm 0.2cm 0cm 1.2cm ,clip, width=\textwidth]{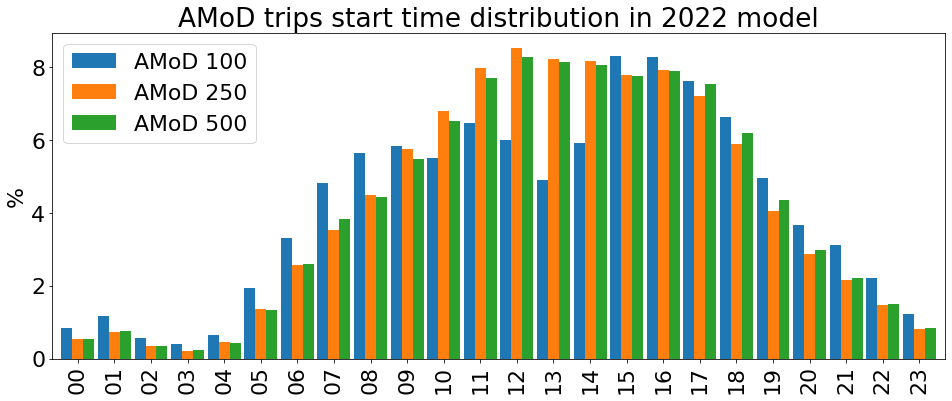}
         \caption{}
         \label{fig:fig14a}
     \end{subfigure}
      \begin{subfigure}[b]{0.48\textwidth}
         \centering
         \includegraphics[trim=0cm 0.2cm 0cm 1.2cm ,clip,width=\textwidth]{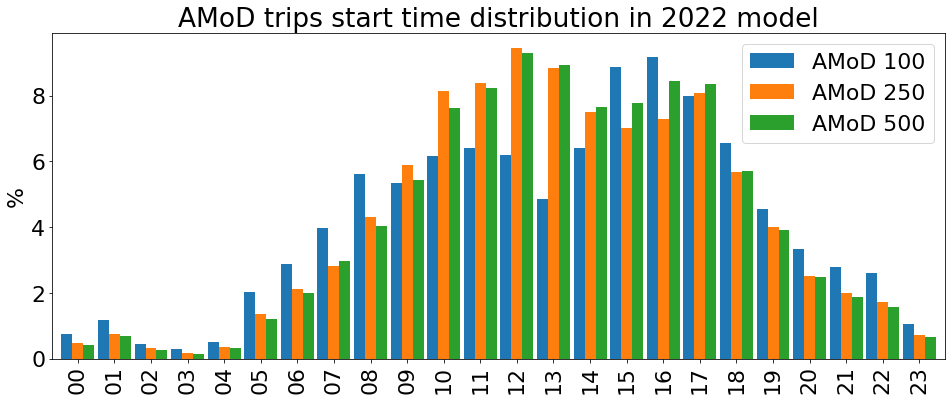}
         \caption{}
         \label{fig:fig14b}
     \end{subfigure}
    \caption{Distributions of AMoD trips’ start time for each scenario: AMoD with 100 vehicles (in blue), AMoD with 250 vehicles (in orange) and AMoD with 500 vehicles (in green), (a) without the induced demand, and (b) with the induced demand.}
    \label{fig:fig14}
\end{figure}

The analysis of AMoD timing (Figure \ref{fig:fig14}) reveals that when the fleet size is limited to 100 vehicles, older adults tend to use AMoD primarily for morning and afternoon trips. As the fleet size increases to 250 or 500 vehicles, older adults rely on AMoD for trips all over the day, since most older adults trips are made with this mode.

When the AMoD service operates with only 100 vehicles, the trips that experience a shift to AMoD are those occurring in the early morning (before 9a.m.), and in the afternoon and evening (after 3p.m., and especially after 8p.m.). This might indicate that the first older adults' trips that switched to AMoD are the ones that older adults have difficulty to perform, e.g. when there is no public transportation or when the traffic flow is slow due to congestion.

When the induced demand is considered, similar observations can be made; the first trips to be shifted to AMoD are those occurring in the early morning, and in the afternoon and evening. As the fleet size increases, the AMoD trips distribution tends to be closer to the older adults trips distribution, i.e. smooth increase in the morning until midday and a smooth decrease after 4p.m.

\begin{figure}[ht]
    \centering
    \begin{subfigure}[b]{0.25\textwidth}
         \centering
         \includegraphics[trim=0cm 0cm 0cm 4cm ,clip, width=\textwidth]{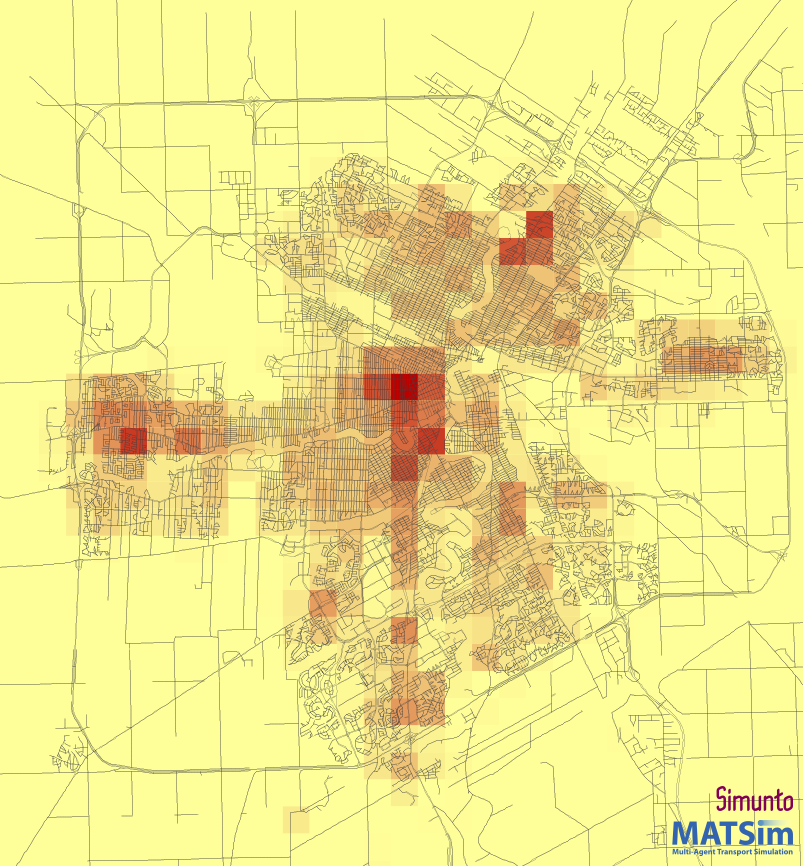}
         \caption{}
         \label{fig:fig13a}
     \end{subfigure}
      \begin{subfigure}[b]{0.25\textwidth}
         \centering
         \includegraphics[trim=0cm 0cm 0cm 4cm ,clip,width=\textwidth]{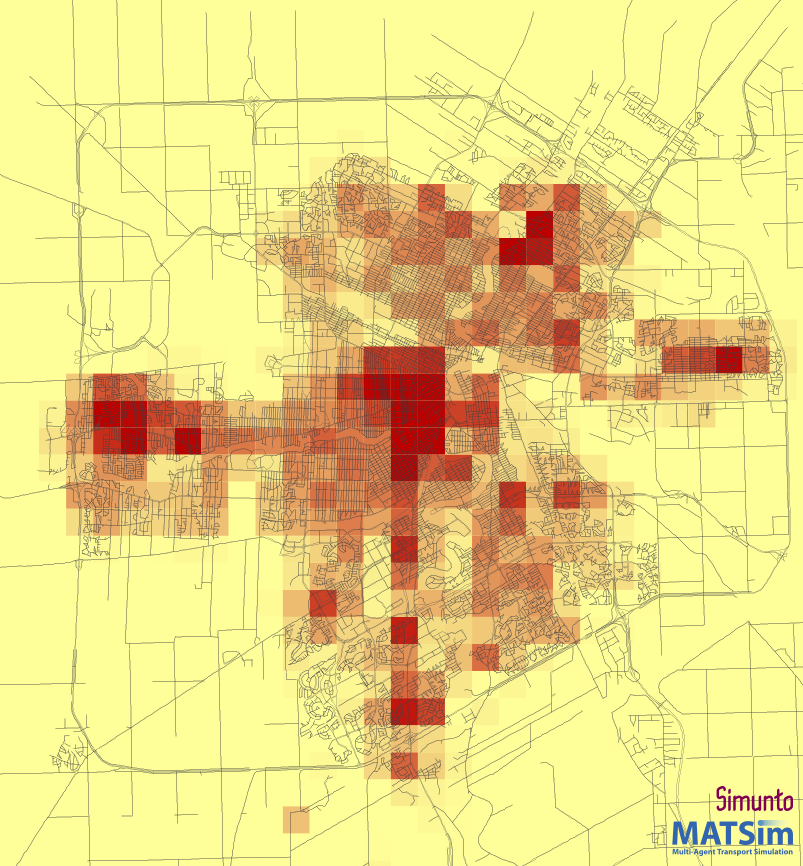}
         \caption{}
         \label{fig:fig13d}
     \end{subfigure}
      \begin{subfigure}[b]{0.25\textwidth}
         \centering
         \includegraphics[trim=0cm 0cm 0cm 4cm ,clip,width=\textwidth]{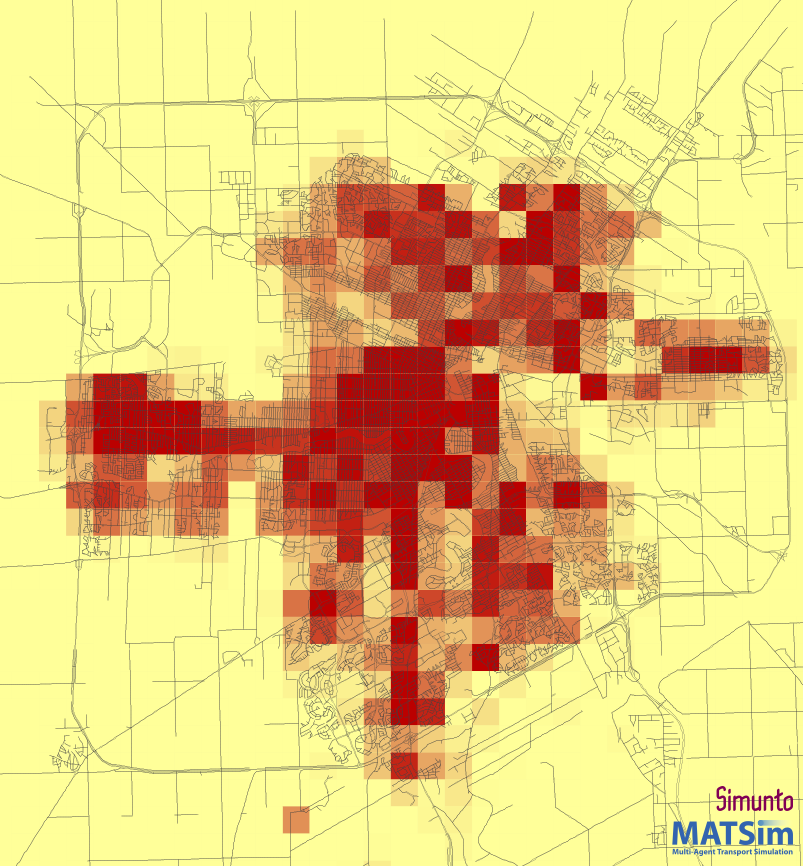}
         \caption{}
         \label{fig:fig13c}
     \end{subfigure}
    \caption{Number of requests by AMoD service zone, with (a) 100, (b) 250 and (c) 500 vehicles. Light yellow is for 0 request and dark red is for 150 requests.}
    \label{fig:fig13}
\end{figure}

Regarding the location of the requests (Figure \ref{fig:fig13}), if the number of vehicles is limited to 100, the requests are concentrated where there is a high density of older adults; in the city centre, in Westwood and Crestview neighbourhood (in the west), and in the River east west neighbourhood (in the north of the city).
As the fleet size increases and more requests can be served, older adults from other zones progressively switch to AMoD. With 500 vehicles, requests are made from all the city. The spatial distributions of the requests are identical for the three repetitions, and whether or not induced demand is considered. These findings may guide the AMoD vehicles spatial distribution in the city for an efficient service.

% Note that the demand has been aggregated to be presented in Figure \ref{fig:fig13} but that the model contains requests locations at the individual scale.

\section{Conclusions}\label{conclu}
% The main conclusions of the study may be presented in a short Conclusions section, which may stand alone or form a subsection of a Discussion or Results and Discussion section.

The Canadian population is ageing, and segments of the older population are experiencing difficulties in meeting their transportation needs. An AMoD service could support their mobility, provided that this new solution is carefully designed in order to meet older adults specific demand.

This paper attempts to estimate older adults' mobility needs by creating an ABM to study older adults demand for an AMoD service in Winnipeg, Canada. This approach incorporates a rich and realistic model of Winnipeg population and its mobility at a fine spatial and temporal scale. The model has been calibrated to obtain a realistic modal split and has been validated against real-world observations. The ABM has been simulated in MATSim, where a simulated AMoD service was added as a new mobility solution for older adults.
Our findings indicate that the model can be used to explore scenarios that can not be explored in real world, such as testing various designs for an AMoD service before its implementation. The simulated demand adapt to the AMoD fleet size, as the real-world demand would adapt to the service quality. The model can thus be used to identify the minimum fleet size needed to absorb the demand and the fleet size with which all requests could be served.
This work shows that an ABM approach is relevant for assessing the AMoD demand of specific population groups in large-scale urbanised areas.

The outputs of the proposed model and simulations shed light on the complex dynamics of AMoD systems and have several implications for transportation planning and policy. First, while increasing fleet size can initially reduce service rejections and improve overall accessibility, it is essential to consider the potential effects of induced demand.
The appropriate fleet size and modal share differ between scenarios with and without induced demand. In this work, we considered the demand resulting from new user groups, i.e. the non-licensed elderly, and from the elderly switching their transportation mode to use AMoD. 
However, part of the demand for an AMoD service could also result from a change in the frequency of trips (e.g., additional trips that were not previously made by users \citep{harb_2018, rahman_evaluation_2020}), a change in the trips length (e.g., longer trips \citep{harb_2018}), shifts in the timing of trips (e.g., more evening trips \citep{harb_2018}), or even changes in the locations of activities (e.g., use of facilities that were not previously accessible). In the longer term, demand could further evolve as a result of changes in where people live; an AMoD service making residential areas with poor public transport links more attractive.

The AMoD implementation should therefore be conducted in hand with infrastructure planning to avoid encouraging undesired urban sprawl and making older adult users dependent on the AMoD service. Future research in this area can delve deeper into understanding the interplay between fleet size, induced demand, and operational efficiency to maximise the benefits of AMoD services.
A limitation of the model is that only 25\% of the Winnipeg population is simulated and used to study the AMoD demand. \cite{kagho_effects_2022} recently showed that if a 25\% population sample could be relevant for analysing some AMoD adoption outputs, a larger sample has to be used to have a complete evaluation of the AMoD design implemented. For example, a 75\% population sample is the minimum required to accurately study the users waiting time, and a 95\% sample is required to estimate the average vehicle occupancy. Future work will focus on calibrating the full population model, validating the mobility using traffic counts (if available) and explore more AMoD scenarios.

Second, if the elderly value the AMoD service as they value the car transportation mode, then the AMoD adoption could be very important. The use of biking, walking and public transit as transportation modes could be reduced to marginal proportions. The use of the car could also be greatly reduced, but would remain the second mode of transport for older adults after the AMoD. Once a certain threshold of AMoD adoption has been reached, almost 30\% of the trips appear to continue to be made by car, even if the AMoD fleet is expanded. It is therefore important to clearly define the objective when setting up an AMoD service; to replace the car for older adults or to serve as a complementary mode of transport for disadvantaged older adults. The size of the fleet and service supply should be defined according to the objective.

Third, when the induced demand is considered, the oldest and female segments of the population, who particularly depend on ride, walking and public transit, could be the most demanding users. This has been observed in real world, where older females with no car access would seem likely to have the greatest number of unmet needs \citep{shergold_mobility_2016}. However, an important limitation of the model is that it does not include a discrete mode choice model. All simulated older adults value the AMoD mode in the same way, while in the real-world, some older adults are more reluctant to use such new service than others, depending on their age, gender, income, and technology acceptance \citep{kersting_2021,hassan_factors_2019,golbabaei_2020}. Combining the proposed model with a discrete mode choice model that integrates older adults heterogeneous acceptance of AMoD would allow to explore the demand evolution more in depth.

Finally, the introduction of AMoD services with the intention of reducing transportation inequalities is a promising initiative. However, for AMoD services to be truly inclusive and fair, their implementation must specifically target the most vulnerable populations. Without particular attention to this aspect, there is a risk that the AMoD service will mainly be adopted by already mobile populations, leaving vulnerable individuals still facing difficulties in getting around.
This was observed in the model with older adults and women starting to use the AMoD significantly more only after the fleet size has been increased.
Older women have shown to be disadvantaged compared to younger men regarding mobility and might be even more disadvantaged with technology-dependent mobility \citep{shergold_mobility_2016}.
To avoid widening the mobility gender gap, it is essential to establish public policies and incentives that facilitate access to AMoD for disadvantaged populations, ensuring that pricing and access to service are tailored to their specific needs. Additionally, training and awareness programs should be implemented to ensure that less technologically savvy individuals are not left behind in the mobility transition.

\bibliography{sn-article}% common bib file


\begin{thebibliography}{74}
\ifx \bisbn   \undefined \def \bisbn  #1{ISBN #1}\fi
\ifx \binits  \undefined \def \binits#1{#1}\fi
\ifx \bauthor  \undefined \def \bauthor#1{#1}\fi
\ifx \batitle  \undefined \def \batitle#1{#1}\fi
\ifx \bjtitle  \undefined \def \bjtitle#1{#1}\fi
\ifx \bvolume  \undefined \def \bvolume#1{\textbf{#1}}\fi
\ifx \byear  \undefined \def \byear#1{#1}\fi
\ifx \bissue  \undefined \def \bissue#1{#1}\fi
\ifx \bfpage  \undefined \def \bfpage#1{#1}\fi
\ifx \blpage  \undefined \def \blpage #1{#1}\fi
\ifx \burl  \undefined \def \burl#1{\textsf{#1}}\fi
\ifx \doiurl  \undefined \def \doiurl#1{\url{https://doi.org/#1}}\fi
\ifx \betal  \undefined \def \betal{\textit{et al.}}\fi
\ifx \binstitute  \undefined \def \binstitute#1{#1}\fi
\ifx \binstitutionaled  \undefined \def \binstitutionaled#1{#1}\fi
\ifx \bctitle  \undefined \def \bctitle#1{#1}\fi
\ifx \beditor  \undefined \def \beditor#1{#1}\fi
\ifx \bpublisher  \undefined \def \bpublisher#1{#1}\fi
\ifx \bbtitle  \undefined \def \bbtitle#1{#1}\fi
\ifx \bedition  \undefined \def \bedition#1{#1}\fi
\ifx \bseriesno  \undefined \def \bseriesno#1{#1}\fi
\ifx \blocation  \undefined \def \blocation#1{#1}\fi
\ifx \bsertitle  \undefined \def \bsertitle#1{#1}\fi
\ifx \bsnm \undefined \def \bsnm#1{#1}\fi
\ifx \bsuffix \undefined \def \bsuffix#1{#1}\fi
\ifx \bparticle \undefined \def \bparticle#1{#1}\fi
\ifx \barticle \undefined \def \barticle#1{#1}\fi
\bibcommenthead
\ifx \bconfdate \undefined \def \bconfdate #1{#1}\fi
\ifx \botherref \undefined \def \botherref #1{#1}\fi
\ifx \url \undefined \def \url#1{\textsf{#1}}\fi
\ifx \bchapter \undefined \def \bchapter#1{#1}\fi
\ifx \bbook \undefined \def \bbook#1{#1}\fi
\ifx \bcomment \undefined \def \bcomment#1{#1}\fi
\ifx \oauthor \undefined \def \oauthor#1{#1}\fi
\ifx \citeauthoryear \undefined \def \citeauthoryear#1{#1}\fi
\ifx \endbibitem  \undefined \def \endbibitem {}\fi
\ifx \bconflocation  \undefined \def \bconflocation#1{#1}\fi
\ifx \arxivurl  \undefined \def \arxivurl#1{\textsf{#1}}\fi
\csname PreBibitemsHook\endcsname

%%% 1
\bibitem[\protect\citeauthoryear{}{}]{afdc_2023}
\begin{botherref}
Alternative Fuels Data Center: Alternative Fueling Station Locator.
\url{https://afdc.energy.gov/stations/}
Accessed 2023-07-18
\end{botherref}
\endbibitem

%%% 2
\bibitem[\protect\citeauthoryear{Boesch et~al.}{2016}]{boesch_2016}
\begin{barticle}
\bauthor{\bsnm{Boesch}, \binits{P.M.}},
\bauthor{\bsnm{Ciari}, \binits{F.}},
\bauthor{\bsnm{Axhausen}, \binits{K.W.}}:
\batitle{Autonomous vehicle fleet sizes required to serve different levels of demand}.
\bjtitle{Transportation Research Record}
\bvolume{2542}(\bissue{1}),
\bfpage{111}--\blpage{119}
(\byear{2016})
\doiurl{10.3141/2542-13}
\end{barticle}
\endbibitem

%%% 3
\bibitem[\protect\citeauthoryear{Ben-Dor et~al.}{2021}]{ben-dor_population_2021}
\begin{barticle}
\bauthor{\bsnm{Ben-Dor}, \binits{G.}},
\bauthor{\bsnm{Ben-Elia}, \binits{E.}},
\bauthor{\bsnm{Benenson}, \binits{I.}}:
\batitle{Population downscaling in multi-agent transportation simulations: A review and case study}.
\bjtitle{Simulation Modelling Practice and Theory}
\bvolume{108},
\bfpage{102233}
(\byear{2021})
\doiurl{10.1016/j.simpat.2020.102233}
\end{barticle}
\endbibitem

%%% 4
\bibitem[\protect\citeauthoryear{Brown et~al.}{2014}]{brown_2014}
\begin{botherref}
\oauthor{\bsnm{Brown}, \binits{A.}},
\oauthor{\bsnm{Gonder}, \binits{J.}},
\oauthor{\bsnm{Repac}, \binits{B.}}:
An analysis of possible energy impacts of automated vehicle.
Road Vehicle Automation, Lecture Notes in Mobility,
137--153
(2014)
\end{botherref}
\endbibitem

%%% 5
\bibitem[\protect\citeauthoryear{Bastarianto et~al.}{2023}]{bastarianto2023agent}
\begin{barticle}
\bauthor{\bsnm{Bastarianto}, \binits{F.F.}},
\bauthor{\bsnm{Hancock}, \binits{T.O.}},
\bauthor{\bsnm{Choudhury}, \binits{C.F.}},
\bauthor{\bsnm{Manley}, \binits{E.}}:
\batitle{Agent-based models in urban transportation: review, challenges, and opportunities}.
\bjtitle{European Transport Research Review}
\bvolume{15}(\bissue{1}),
\bfpage{1}--\blpage{20}
(\byear{2023})
\end{barticle}
\endbibitem

%%% 6
\bibitem[\protect\citeauthoryear{Bischoff et~al.}{2019}]{Bischoff2019}
\begin{bchapter}
\bauthor{\bsnm{Bischoff}, \binits{J.}},
\bauthor{\bsnm{Márquez-Fernández}, \binits{F.J.}},
\bauthor{\bsnm{Domingues-Olavarría}, \binits{G.}},
\bauthor{\bsnm{Maciejewski}, \binits{M.}},
\bauthor{\bsnm{Nagel}, \binits{K.}}:
\bctitle{Impacts of vehicle fleet electrification in sweden – a simulation-based assessment of long-distance trips}.
In: \bbtitle{The 6th International Conference on Models and Technologies for Intelligent Transportation Systems (MT-ITS)},
pp. \bfpage{1}--\blpage{7}
(\byear{2019}).
\doiurl{10.1109/MTITS.2019.8883384}
\end{bchapter}
\endbibitem

%%% 7
\bibitem[\protect\citeauthoryear{Bischoff et~al.}{2017}]{Bischoff2017}
\begin{bchapter}
\bauthor{\bsnm{Bischoff}, \binits{J.}},
\bauthor{\bsnm{Maciejewski}, \binits{M.}},
\bauthor{\bsnm{Nagel}, \binits{K.}}:
\bctitle{City-wide shared taxis: A simulation study in berlin}.
In: \bbtitle{2017 IEEE 20th International Conference on Intelligent Transportation Systems (ITSC)},
pp. \bfpage{275}--\blpage{280}
(\byear{2017}).
\doiurl{10.1109/ITSC.2017.8317926}
\end{bchapter}
\endbibitem

%%% 8
\bibitem[\protect\citeauthoryear{Bosina and Weidmann}{2017}]{bosina_estimating_2017}
\begin{barticle}
\bauthor{\bsnm{Bosina}, \binits{E.}},
\bauthor{\bsnm{Weidmann}, \binits{U.}}:
\batitle{Estimating pedestrian speed using aggregated literature data}.
\bjtitle{Physica A: Statistical Mechanics and its Applications}
\bvolume{468},
\bfpage{1}--\blpage{29}
(\byear{2017})
\doiurl{10.1016/j.physa.2016.09.044}
\end{barticle}
\endbibitem

%%% 9
\bibitem[\protect\citeauthoryear{Currie and Fournier}{2020}]{currie_why_2020}
\begin{barticle}
\bauthor{\bsnm{Currie}, \binits{G.}},
\bauthor{\bsnm{Fournier}, \binits{N.}}:
\batitle{{Why most {DRT}/Micro-Transits fail – What the survivors tell us about progress}}.
\bjtitle{Research in Transportation Economics}
\bvolume{83},
\bfpage{100895}
(\byear{2020})
\doiurl{10.1016/j.retrec.2020.100895}
\end{barticle}
\endbibitem

%%% 10
\bibitem[\protect\citeauthoryear{Franco et~al.}{2020}]{franco_demand_2020}
\begin{barticle}
\bauthor{\bsnm{Franco}, \binits{P.}},
\bauthor{\bsnm{Johnston}, \binits{R.}},
\bauthor{\bsnm{{McCormick}}, \binits{E.}}:
\batitle{Demand responsive transport: Generation of activity patterns from mobile phone network data to support the operation of new mobility services}.
\bjtitle{Transportation Research Part A: Policy and Practice}
\bvolume{131},
\bfpage{244}--\blpage{266}
(\byear{2020})
\doiurl{10.1016/j.tra.2019.09.038}
\end{barticle}
\endbibitem

%%% 11
\bibitem[\protect\citeauthoryear{Fagnant et~al.}{2016}]{fagnant_2016}
\begin{barticle}
\bauthor{\bsnm{Fagnant}, \binits{D.J.}},
\bauthor{\bsnm{Kockelman}, \binits{K.M.}},
\bauthor{\bsnm{Bansal}, \binits{P.}}:
\batitle{{Operations of shared autonomous vehicle fleet for Austin, Texas, market}}.
\bjtitle{Transportation Research Record}
\bvolume{2563}(\bissue{1}),
\bfpage{98}--\blpage{106}
(\byear{2016})
\doiurl{10.3141/2536-12}
\end{barticle}
\endbibitem

%%% 12
\bibitem[\protect\citeauthoryear{Fatima et~al.}{2020}]{fatima_elderly_2020}
\begin{barticle}
\bauthor{\bsnm{Fatima}, \binits{K.}},
\bauthor{\bsnm{Moridpour}, \binits{S.}},
\bauthor{\bsnm{De~Gruyter}, \binits{C.}},
\bauthor{\bsnm{Saghapour}, \binits{T.}}:
\batitle{Elderly sustainable mobility: Scientific paper review}.
\bjtitle{Sustainability}
\bvolume{12}(\bissue{18}),
\bfpage{7319}
(\byear{2020})
\doiurl{10.3390/su12187319}
\end{barticle}
\endbibitem

%%% 13
\bibitem[\protect\citeauthoryear{Faber and van Lierop}{2020}]{faber_how_2020}
\begin{barticle}
\bauthor{\bsnm{Faber}, \binits{K.}},
\bauthor{\bsnm{Lierop}, \binits{D.}}:
\batitle{{How will older adults use automated vehicles? Assessing the role of {AVs} in overcoming perceived mobility barriers}}.
\bjtitle{Transportation Research Part A: Policy and Practice}
\bvolume{133},
\bfpage{353}--\blpage{363}
(\byear{2020})
\doiurl{10.1016/j.tra.2020.01.022}
\end{barticle}
\endbibitem

%%% 14
\bibitem[\protect\citeauthoryear{Graur et~al.}{2021}]{hermes2021}
\begin{bchapter}
\bauthor{\bsnm{Graur}, \binits{D.}},
\bauthor{\bsnm{Bruno}, \binits{R.}},
\bauthor{\bsnm{Bischoff}, \binits{J.}},
\bauthor{\bsnm{Rieser}, \binits{M.}},
\bauthor{\bsnm{Scherr}, \binits{W.}},
\bauthor{\bsnm{Hoefler}, \binits{T.}},
\bauthor{\bsnm{Alonso}, \binits{G.}}:
\bctitle{Hermes: Enabling efficient large-scale simulation in matsim}.
In: \bbtitle{The 12th International Conference on Ambient Systems, Networks and Technologies (ANT) / The 4th International Conference on Emerging Data and Industry 4.0 (EDI40) / Affiliated Workshops},
vol. \bseriesno{184},
pp. \bfpage{635}--\blpage{641}
(\byear{2021}).
\doiurl{10.1016/j.procs.2021.03.079} .
\burl{https://www.sciencedirect.com/science/article/pii/S1877050921007158}
\end{bchapter}
\endbibitem

%%% 15
\bibitem[\protect\citeauthoryear{Golbabaei et~al.}{2020}]{golbabaei_2020}
\begin{barticle}
\bauthor{\bsnm{Golbabaei}, \binits{F.}},
\bauthor{\bsnm{Yigitcanlar}, \binits{T.}},
\bauthor{\bsnm{Paz}, \binits{A.}},
\bauthor{\bsnm{Bunker}, \binits{J.}}:
\batitle{Individual predictors of autonomous vehicle public acceptance and intention to use: A systematic review of the literature}.
\bjtitle{Journal of Open Innovation: Technology, Market, and Complexity}
\bvolume{6}(\bissue{4}),
\bfpage{106}
(\byear{2020})
\doiurl{10.3390/joitmc6040106}
\end{barticle}
\endbibitem

%%% 16
\bibitem[\protect\citeauthoryear{Haustein}{2012}]{haustein_mobility_2012}
\begin{barticle}
\bauthor{\bsnm{Haustein}, \binits{S.}}:
\batitle{Mobility behavior of the elderly: an attitude-based segmentation approach for a heterogeneous target group}.
\bjtitle{Transportation}
\bvolume{39}(\bissue{6}),
\bfpage{1079}--\blpage{1103}
(\byear{2012})
\doiurl{10.1007/s11116-011-9380-7} .
Accessed 2022-03-07
\end{barticle}
\endbibitem

%%% 17
\bibitem[\protect\citeauthoryear{Hörl and Balac}{2021}]{horl_synthetic_2021}
\begin{barticle}
\bauthor{\bsnm{Hörl}, \binits{S.}},
\bauthor{\bsnm{Balac}, \binits{M.}}:
\batitle{{Synthetic population and travel demand for Paris and Île-de-France based on open and publicly available data}}.
\bjtitle{Transportation Research Part C: Emerging Technologies}
\bvolume{130},
\bfpage{103291}
(\byear{2021})
\doiurl{10.1016/j.trc.2021.103291}
\end{barticle}
\endbibitem

%%% 18
\bibitem[\protect\citeauthoryear{Hörl et~al.}{2019}]{horl_2019}
\begin{bchapter}
\bauthor{\bsnm{Hörl}, \binits{S.}},
\bauthor{\bsnm{Balac}, \binits{M.}},
\bauthor{\bsnm{Axhausen}, \binits{K.}}:
\bctitle{{Dynamic demand estimation for an AMoD system in Paris}}.
In: \bbtitle{Intelligent Vehicles Symposium 2019},
pp. \bfpage{260}--\blpage{266}
(\byear{2019}).
\doiurl{10.1109/IVS.2019.8814051}
\end{bchapter}
\endbibitem

%%% 19
\bibitem[\protect\citeauthoryear{Harper et~al.}{2016}]{harper_2016}
\begin{botherref}
\oauthor{\bsnm{Harper}, \binits{C.D.}},
\oauthor{\bsnm{Chris~T.}, \binits{H.}},
\oauthor{\bsnm{Sonia}, \binits{M.}},
\oauthor{\bsnm{Constantine}, \binits{S.}}:
Estimating potential increases in travel with autonomous vehicles for the non-driving, elderly and people with travel-restrictive medical conditions.
Transportation Research Part C: Emerging Technologie,
1--9
(2016)
\doiurl{10.1016/j.trc.2016.09.003}
\end{botherref}
\endbibitem

%%% 20
\bibitem[\protect\citeauthoryear{Hassan et~al.}{2019}]{hassan_factors_2019}
\begin{barticle}
\bauthor{\bsnm{Hassan}, \binits{H.M.}},
\bauthor{\bsnm{Ferguson}, \binits{M.R.}},
\bauthor{\bsnm{Razavi}, \binits{S.}},
\bauthor{\bsnm{Vrkljan}, \binits{B.}}:
\batitle{Factors that influence older {Canadians}’ preferences for using autonomous vehicle technology: A structural equation analysis}.
\bjtitle{Transportation Research Record: Journal of the Transportation Research Board}
\bvolume{2673}(\bissue{1}),
\bfpage{469}--\blpage{480}
(\byear{2019})
\doiurl{10.1177/0361198118822281}
\end{barticle}
\endbibitem

%%% 21
\bibitem[\protect\citeauthoryear{Hogeveen et~al.}{2021}]{hogeveen_2021}
\begin{botherref}
\oauthor{\bsnm{Hogeveen}, \binits{P.}},
\oauthor{\bsnm{Maarten}, \binits{S.}},
\oauthor{\bsnm{Geert}, \binits{V.}},
\oauthor{\bsnm{Auke}, \binits{H.}}:
Quantifying the fleet composition at full adoption of shared autonomous electric vehicles: An agent-based approach.
The Open Transportation Journal,
47--60
(2021)
\doiurl{10.2174/1874447802115010047}
\end{botherref}
\endbibitem

%%% 22
\bibitem[\protect\citeauthoryear{Horni et~al.}{2016}]{horni2016introducing}
\begin{bchapter}
\bauthor{\bsnm{Horni}, \binits{A.}},
\bauthor{\bsnm{Nagel}, \binits{K.}},
\bauthor{\bsnm{Axhausen}, \binits{K.W.}}:
\bctitle{{Introducing MATSim}}.
In: \bbtitle{The Multi-agent Transport Simulation {MATSim}},
pp. \bfpage{3}--\blpage{7}.
\bpublisher{Ubiquity Press}, \blocation{???}
(\byear{2016})
\end{bchapter}
\endbibitem

%%% 23
\bibitem[\protect\citeauthoryear{Harb et~al.}{2018}]{harb_2018}
\begin{botherref}
\oauthor{\bsnm{Harb}, \binits{M.}},
\oauthor{\bsnm{Xiao}, \binits{Y.}},
\oauthor{\bsnm{Circella}, \binits{G.}},
\oauthor{\bsnm{Mokhtarian}, \binits{P.L.}},
\oauthor{\bsnm{Walker}, \binits{J.L.}}:
Projecting travelers into a world of self-driving vehicles: Estimating travel behavior implications via a naturalistic experiment.
Transportation,
1671--85
(2018)
\doiurl{10.1007/s11116-018-9937-9}
\end{botherref}
\endbibitem

%%% 24
\bibitem[\protect\citeauthoryear{iTRANS Project~Team}{2009}]{wats_2007}
\begin{botherref}
\oauthor{\bsnm{Project~Team}}:
2007 winnipeg area travel survey results - final report.
Technical report,
iTRANS Consulting Inc
(2009).
\url{https://legacy.winnipeg.ca/publicworks/transportation/pdf/transportationMasterPlan/WATS-Final-Report-July2007.pdf}
\end{botherref}
\endbibitem

%%% 25
\bibitem[\protect\citeauthoryear{Kouros et~al.}{2013}]{mohammadian_2013}
\begin{botherref}
\oauthor{\bsnm{Kouros}, \binits{M.}},
\oauthor{\bsnm{Behzad}, \binits{K.}},
\oauthor{\bsnm{Zahra}, \binits{P.}},
\oauthor{\bsnm{Martina}, \binits{F.}}:
Modeling seniors’ activity-travel data.
Technical report,
Illinois Center for Transportation
(2013)
\end{botherref}
\endbibitem

%%% 26
\bibitem[\protect\citeauthoryear{Kagho et~al.}{2021}]{kagho_2021}
\begin{barticle}
\bauthor{\bsnm{Kagho}, \binits{G.O.}},
\bauthor{\bsnm{Hensle}, \binits{D.}},
\bauthor{\bsnm{Balac}, \binits{M.}},
\bauthor{\bsnm{Freedman}, \binits{J.}},
\bauthor{\bsnm{Twumasi-Boakye}, \binits{R.}},
\bauthor{\bsnm{Broaddus}, \binits{A.}},
\bauthor{\bsnm{Fishelson}, \binits{J.}},
\bauthor{\bsnm{Axhausen}, \binits{K.W.}}:
\batitle{Demand responsive transit simulation of {Wayne county, Michigan}}.
\bjtitle{Transportation Research Record}
\bvolume{2675}(\bissue{12}),
\bfpage{702}--\blpage{716}
(\byear{2021})
\doiurl{10.1177/03611981211031221}
\end{barticle}
\endbibitem

%%% 27
\bibitem[\protect\citeauthoryear{Kersting et~al.}{2021}]{kersting_2021}
\begin{barticle}
\bauthor{\bsnm{Kersting}, \binits{M.}},
\bauthor{\bsnm{Kallbach}, \binits{F.}},
\bauthor{\bsnm{Schlüter}, \binits{J.C.}}:
\batitle{{For the young and old alike – An analysis of the determinants of seniors’ satisfaction with the true door-to-door {DRT} system {EcoBus} in rural Germany}}.
\bjtitle{Journal of Transport Geography}
\bvolume{96},
\bfpage{103173}
(\byear{2021})
\doiurl{10.1016/j.jtrangeo.2021.103173}
\end{barticle}
\endbibitem

%%% 28
\bibitem[\protect\citeauthoryear{Kaddoura et~al.}{2020}]{kaddoura_impact_2020}
\begin{barticle}
\bauthor{\bsnm{Kaddoura}, \binits{I.}},
\bauthor{\bsnm{Leich}, \binits{G.}},
\bauthor{\bsnm{Nagel}, \binits{K.}}:
\batitle{The impact of pricing and service area design on the modal shift towards demand responsive transit}.
\bjtitle{Procedia Computer Science}
\bvolume{170},
\bfpage{807}--\blpage{812}
(\byear{2020})
\doiurl{10.1016/j.procs.2020.03.152}
\end{barticle}
\endbibitem

%%% 29
\bibitem[\protect\citeauthoryear{Kovacs et~al.}{2020}]{kovacs2020aged}
\begin{barticle}
\bauthor{\bsnm{Kovacs}, \binits{F.S.}},
\bauthor{\bsnm{McLeod}, \binits{S.}},
\bauthor{\bsnm{Curtis}, \binits{C.}}:
\batitle{Aged mobility in the era of transportation disruption: Will autonomous vehicles address impediments to the mobility of ageing populations?}
\bjtitle{Travel behaviour and society}
\bvolume{20},
\bfpage{122}--\blpage{132}
(\byear{2020})
\end{barticle}
\endbibitem

%%% 30
\bibitem[\protect\citeauthoryear{Kagho et~al.}{2022}]{kagho_effects_2022}
\begin{barticle}
\bauthor{\bsnm{Kagho}, \binits{G.O.}},
\bauthor{\bsnm{Meli}, \binits{J.}},
\bauthor{\bsnm{Walser}, \binits{D.}},
\bauthor{\bsnm{Balac}, \binits{M.}}:
\batitle{Effects of population sampling on agent-based transport simulation of on-demand services}.
\bjtitle{Procedia Computer Science}
\bvolume{201},
\bfpage{305}--\blpage{312}
(\byear{2022})
\doiurl{10.1016/j.procs.2022.03.041}
\end{barticle}
\endbibitem

%%% 31
\bibitem[\protect\citeauthoryear{Li et~al.}{2012}]{li_population_2012}
\begin{barticle}
\bauthor{\bsnm{Li}, \binits{H.}},
\bauthor{\bsnm{Raeside}, \binits{R.}},
\bauthor{\bsnm{Chen}, \binits{T.}},
\bauthor{\bsnm{{McQuaid}}, \binits{R.W.}}:
\batitle{Population ageing, gender and the transportation system}.
\bjtitle{Research in Transportation Economics}
\bvolume{34}(\bissue{1}),
\bfpage{39}--\blpage{47}
(\byear{2012})
\doiurl{10.1016/j.retrec.2011.12.007}
\end{barticle}
\endbibitem

%%% 32
\bibitem[\protect\citeauthoryear{Luiu and Tight}{2021}]{luiu_travel_2021}
\begin{barticle}
\bauthor{\bsnm{Luiu}, \binits{C.}},
\bauthor{\bsnm{Tight}, \binits{M.}}:
\batitle{Travel difficulties and barriers during later life: Evidence from the national travel survey in {England}}.
\bjtitle{Journal of Transport Geography}
\bvolume{91},
\bfpage{102973}
(\byear{2021})
\doiurl{10.1016/j.jtrangeo.2021.102973}
\end{barticle}
\endbibitem

%%% 33
\bibitem[\protect\citeauthoryear{Müller and Axhausen}{2014}]{muller_2014}
\begin{barticle}
\bauthor{\bsnm{Müller}, \binits{K.}},
\bauthor{\bsnm{Axhausen}, \binits{K.W.}}:
\batitle{Using survey calibration and statistical matching to reweight and distribute activity schedules}.
\bjtitle{Transportation Research Record}
\bvolume{2429}(\bissue{1}),
\bfpage{157}--\blpage{167}
(\byear{2014})
\doiurl{10.3141/2429-17}
\end{barticle}
\endbibitem

%%% 34
\bibitem[\protect\citeauthoryear{{Manitoba Public Insurance}}{2020}]{mpi}
\begin{botherref}
\oauthor{\bsnm{{Manitoba Public Insurance}}}:
{2020 Traffic collision statistics report}.
Available at \url{https://www.mpi.mb.ca/Pages/traffic-collision-report.aspx}.
Accessed: 2022-09
(2020)
\end{botherref}
\endbibitem

%%% 35
\bibitem[\protect\citeauthoryear{{Microsoft and Statistics Canada}}{2019}]{microsoft}
\begin{botherref}
\oauthor{\bsnm{{Microsoft and Statistics Canada}}}:
{Canadian Building Footprints}.
Available at \url{https://github.com/Microsoft/CanadianBuildingFootprints}.
Accessed: 2022-12
(2019)
\end{botherref}
\endbibitem

%%% 36
\bibitem[\protect\citeauthoryear{M{\"u}ller et~al.}{2021}]{muller_2021}
\begin{bchapter}
\bauthor{\bsnm{M{\"u}ller}, \binits{J.}},
\bauthor{\bsnm{Straub}, \binits{M.}},
\bauthor{\bsnm{Naqvi}, \binits{A.}},
\bauthor{\bsnm{Richter}, \binits{G.}},
\bauthor{\bsnm{Peer}, \binits{S.}},
\bauthor{\bsnm{Rudloff}, \binits{C.}}:
\bctitle{{MATSim} model {Vienna}: Analyzing the socioeconomic impacts for different fleet sizes and pricing schemes of shared autonomous electric vehicles}.
In: \bbtitle{Transportation Research Board 100th Annual Meeting 2021}
(\byear{2021})
\end{bchapter}
\endbibitem

%%% 37
\bibitem[\protect\citeauthoryear{Militão and Tirachini}{2021}]{militao_optimal_2021}
\begin{barticle}
\bauthor{\bsnm{Militão}, \binits{A.M.}},
\bauthor{\bsnm{Tirachini}, \binits{A.}}:
\batitle{Optimal fleet size for a shared demand-responsive transport system with human-driven vs automated vehicles: A total cost minimization approach}.
\bjtitle{Transportation Research Part A: Policy and Practice}
\bvolume{151},
\bfpage{52}--\blpage{80}
(\byear{2021})
\doiurl{10.1016/j.tra.2021.07.004}
\end{barticle}
\endbibitem

%%% 38
\bibitem[\protect\citeauthoryear{Namazi-Rad et~al.}{2017}]{namazirad_2017}
\begin{barticle}
\bauthor{\bsnm{Namazi-Rad}, \binits{M.-R.}},
\bauthor{\bsnm{Tanton}, \binits{R.}},
\bauthor{\bsnm{Steel}, \binits{D.}},
\bauthor{\bsnm{Mokhtarian}, \binits{P.}},
\bauthor{\bsnm{Das}, \binits{S.}}:
\batitle{An unconstrained statistical matching algorithm for combining individual and household level geo-specific census and survey data}.
\bjtitle{Computers, Environment and Urban Systems}
\bvolume{63},
\bfpage{3}--\blpage{14}
(\byear{2017})
\doiurl{10.1016/j.compenvurbsys.2016.11.003}
\end{barticle}
\endbibitem

%%% 39
\bibitem[\protect\citeauthoryear{Newbold et~al.}{2005}]{newbold_travel_2005}
\begin{barticle}
\bauthor{\bsnm{Newbold}, \binits{K.B.}},
\bauthor{\bsnm{Scott}, \binits{D.M.}},
\bauthor{\bsnm{Spinney}, \binits{J.E.L.}},
\bauthor{\bsnm{Kanaroglou}, \binits{P.}},
\bauthor{\bsnm{Páez}, \binits{A.}}:
\batitle{{Travel behavior within Canada’s older population: a cohort analysis}}.
\bjtitle{Journal of Transport Geography}
\bvolume{13}(\bissue{4}),
\bfpage{340}--\blpage{351}
(\byear{2005})
\doiurl{10.1016/j.jtrangeo.2004.07.007}
\end{barticle}
\endbibitem

%%% 40
\bibitem[\protect\citeauthoryear{O'Hern and Oxley}{2015}]{ohern_understanding_2015}
\begin{barticle}
\bauthor{\bsnm{O'Hern}, \binits{S.}},
\bauthor{\bsnm{Oxley}, \binits{J.}}:
\batitle{Understanding travel patterns to support safe active transport for older adults}.
\bjtitle{Journal of Transport \& Health}
\bvolume{2}(\bissue{1}),
\bfpage{79}--\blpage{85}
(\byear{2015})
\doiurl{10.1016/j.jth.2014.09.016}
\end{barticle}
\endbibitem

%%% 41
\bibitem[\protect\citeauthoryear{{OpenStreetMap contributors}}{2017}]{osm}
\begin{botherref}
\oauthor{\bsnm{{OpenStreetMap contributors}}}:
{OpenStreetMap}.
Available at \url{ https://www.openstreetmap.org }.
Accessed: 2022-05
(2017)
\end{botherref}
\endbibitem

%%% 42
\bibitem[\protect\citeauthoryear{O'fallon and Sullivan}{2009}]{ofallon_2009}
\begin{botherref}
\oauthor{\bsnm{O'fallon}, \binits{C.}},
\oauthor{\bsnm{Sullivan}, \binits{C.}}:
Trends in older people’s travel patterns: Analysing changes in older {New Zealanders}’ travel patterns using the ongoing {New Zealand} household travel survey (research report rr 369).
Technical report,
NZ Transport Agency
(2009)
\end{botherref}
\endbibitem

%%% 43
\bibitem[\protect\citeauthoryear{Oh et~al.}{2020}]{oh_assessing_2020}
\begin{barticle}
\bauthor{\bsnm{Oh}, \binits{S.}},
\bauthor{\bsnm{Seshadri}, \binits{R.}},
\bauthor{\bsnm{Azevedo}, \binits{C.L.}},
\bauthor{\bsnm{Kumar}, \binits{N.}},
\bauthor{\bsnm{Basak}, \binits{K.}},
\bauthor{\bsnm{Ben-Akiva}, \binits{M.}}:
\batitle{{Assessing the impacts of automated mobility-on-demand through agent-based simulation: A study of Singapore}}.
\bjtitle{Transportation Research Part A: Policy and Practice}
\bvolume{138},
\bfpage{367}--\blpage{388}
(\byear{2020})
\doiurl{10.1016/j.tra.2020.06.004}
\end{barticle}
\endbibitem

%%% 44
\bibitem[\protect\citeauthoryear{Parminder et~al.}{2018}]{clsa}
\begin{botherref}
\oauthor{\bsnm{Parminder}, \binits{R.}},
\oauthor{\bsnm{Christina}, \binits{W.}},
\oauthor{\bsnm{Susan}, \binits{K.}},
\oauthor{\bsnm{Lauren}, \binits{G.}}:
The {Canadian} longitudinal study on aging ({CLSA}) report on health and aging in {Canada} : Findings from baseline data collection 2010-2015.
Technical report,
The Canadian Longitudinal Study on Aging
(2018).
\url{https://ifa.ngo/wp-content/uploads/2018/12/clsa_report_en_final_web.pdf}
\end{botherref}
\endbibitem

%%% 45
\bibitem[\protect\citeauthoryear{Prédhumeau and Manley}{2023a}]{predhumeau_2023}
\begin{botherref}
\oauthor{\bsnm{Prédhumeau}, \binits{M.}},
\oauthor{\bsnm{Manley}, \binits{E.}}:
{A synthetic population for agent-based modelling in Canada}.
Scientific Data
\textbf{10}
(2023)
\doiurl{10.1038/s41597-023-02030-4}
\end{botherref}
\endbibitem

%%% 46
\bibitem[\protect\citeauthoryear{Prédhumeau and Manley}{2023b}]{predhumeau_cupum_2023}
\begin{bchapter}
\bauthor{\bsnm{Prédhumeau}, \binits{M.}},
\bauthor{\bsnm{Manley}, \binits{E.}}:
\bctitle{{Building an environment for spatial modelling of Canadian cities using open data: A replicable workflow applied to Winnipeg}}.
In: \bbtitle{The 18th International Conference on Computational Urban Planning and Urban Management}
(\byear{2023}).
\doiurl{10.17605/OSF.IO/6YR5V}
\end{bchapter}
\endbibitem

%%% 47
\bibitem[\protect\citeauthoryear{Poletti}{2021}]{pt2matsim}
\begin{botherref}
\oauthor{\bsnm{Poletti}, \binits{F.}}:
{PT2MATSim v. 22.11}.
Available at \url{https://github.com/matsim-org/pt2matsim}.
Accessed: 2022-10
(2021)
\end{botherref}
\endbibitem

%%% 48
\bibitem[\protect\citeauthoryear{Rahman et~al.}{2020}]{rahman_evaluation_2020}
\begin{barticle}
\bauthor{\bsnm{Rahman}, \binits{M.M.}},
\bauthor{\bsnm{Deb}, \binits{S.}},
\bauthor{\bsnm{Strawderman}, \binits{L.}},
\bauthor{\bsnm{Smith}, \binits{B.}},
\bauthor{\bsnm{Burch}, \binits{R.}}:
\batitle{Evaluation of transportation alternatives for aging population in the era of self-driving vehicles}.
\bjtitle{{IATSS} Research}
\bvolume{44}(\bissue{1}),
\bfpage{30}--\blpage{35}
(\byear{2020})
\doiurl{10.1016/j.iatssr.2019.05.004}
\end{barticle}
\endbibitem

%%% 49
\bibitem[\protect\citeauthoryear{Shahid et~al.}{2009}]{shadid_2009}
\begin{botherref}
\oauthor{\bsnm{Shahid}, \binits{R.}},
\oauthor{\bsnm{Bertazzon}, \binits{S.}},
\oauthor{\bsnm{Knudtson}, \binits{M.L.}},
\oauthor{\bsnm{Ghali}, \binits{W.A.}}:
Comparison of distance measures in spatial analytical modeling for health service planning.
BMC Health Services Research
\textbf{9}
(2009)
\doiurl{10.1186/1472-6963-9-200}
\end{botherref}
\endbibitem

%%% 50
\bibitem[\protect\citeauthoryear{Schlüter et~al.}{2021}]{schluter_impact_2021}
\begin{barticle}
\bauthor{\bsnm{Schlüter}, \binits{J.}},
\bauthor{\bsnm{Bossert}, \binits{A.}},
\bauthor{\bsnm{Rössy}, \binits{P.}},
\bauthor{\bsnm{Kersting}, \binits{M.}}:
\batitle{Impact assessment of autonomous demand responsive transport as a link between urban and rural areas}.
\bjtitle{Research in Transportation Business \& Management}
\bvolume{39},
\bfpage{100613}
(\byear{2021})
\doiurl{10.1016/j.rtbm.2020.100613}
\end{barticle}
\endbibitem

%%% 51
\bibitem[\protect\citeauthoryear{Smith et~al.}{2017}]{smith_2017}
\begin{barticle}
\bauthor{\bsnm{Smith}, \binits{A.}},
\bauthor{\bsnm{Lovelace}, \binits{R.}},
\bauthor{\bsnm{Birkin}, \binits{M.}}:
\batitle{{Population Synthesis with Quasirandom Integer Sampling}}.
\bjtitle{Journal of Artificial Societies and Social Simulation}
\bvolume{20}(\bissue{4}),
\bfpage{14}
(\byear{2017})
\doiurl{10.18564/jasss.3550}
\end{barticle}
\endbibitem

%%% 52
\bibitem[\protect\citeauthoryear{Schm{\"o}cker et~al.}{2005}]{schmocker2005estimating}
\begin{barticle}
\bauthor{\bsnm{Schm{\"o}cker}, \binits{J.-D.}},
\bauthor{\bsnm{Quddus}, \binits{M.A.}},
\bauthor{\bsnm{Noland}, \binits{R.B.}},
\bauthor{\bsnm{Bell}, \binits{M.G.}}:
\batitle{{Estimating trip generation of elderly and disabled people: Analysis of London data}}.
\bjtitle{Transportation research record}
\bvolume{1924}(\bissue{1}),
\bfpage{9}--\blpage{18}
(\byear{2005})
\end{barticle}
\endbibitem

%%% 53
\bibitem[\protect\citeauthoryear{Salman et~al.}{2023}]{salman_2023}
\begin{barticle}
\bauthor{\bsnm{Salman}, \binits{F.}},
\bauthor{\bsnm{Sisiopiku}, \binits{V.P.}},
\bauthor{\bsnm{Khalil}, \binits{J.}},
\bauthor{\bsnm{Yang}, \binits{W.}},
\bauthor{\bsnm{Yan}, \binits{D.}}:
\batitle{Operational impacts of on-demand ride-pooling service options in {Birmingham, AL}}.
\bjtitle{Future Transportation}
\bvolume{3}(\bissue{2}),
\bfpage{519}--\blpage{534}
(\byear{2023})
\doiurl{10.3390/futuretransp3020030}
\end{barticle}
\endbibitem

%%% 54
\bibitem[\protect\citeauthoryear{{Statistics Canada}}{}]{census_2006}
\begin{botherref}
\oauthor{\bsnm{{Statistics Canada}}}:
Mode of transportation for the employed labour force 15 Years and over having a usual place of work , 2006 Census - 20\% Sample Data.
\url{https://www12.statcan.gc.ca/census-recensement/2006/dp-pd/tbt/Rp-eng.cfm?TABID=2&LANG=E&APATH=3&DETAIL=0&DIM=0&FL=A&FREE=0&GC=0&GID=858718&GK=0&GRP=1&PID=90658&PRID=0&PTYPE=88971,97154&S=0&SHOWALL=0&SUB=763&Temporal=2006&THEME=76&VID=0&VNAMEE=&VNAMEF=&D1=0&D2=0&D3=0&D4=0&D5=0&D6=0}
\end{botherref}
\endbibitem

%%% 55
\bibitem[\protect\citeauthoryear{{Statistics Canada}}{2009}]{cchs_2009}
\begin{botherref}
\oauthor{\bsnm{{Statistics Canada}}}:
{Canadian Community Health Survey – Healthy Aging - Most common form of transportation, by sex, marital status, number of chronic conditions, and health perception, seniors aged 65 and older, Canada}.
Technical report,
Statistics Canada
(2009)
\end{botherref}
\endbibitem

%%% 56
\bibitem[\protect\citeauthoryear{{Statistics Canada}}{2015}]{tus_2015}
\begin{botherref}
\oauthor{\bsnm{{Statistics Canada}}}:
Time Use Survey
(2015).
\url{https://www150.statcan.gc.ca/n1/pub/89-658-x/89-658-x2017001-eng.htm}
\end{botherref}
\endbibitem

%%% 57
\bibitem[\protect\citeauthoryear{{Statistics Canada}}{2016a}]{census_2016}
\begin{botherref}
\oauthor{\bsnm{{Statistics Canada}}}:
{2016 Census Profile for Canada, provinces, territories, CDs, CSDs and DAs - Catalogue no. 98-401-X2016044}.
Available at \url{https://www150.statcan.gc.ca/n1/en/catalogue/98-316-X2016001}.
Accessed: 2022-08
(2016)
\end{botherref}
\endbibitem

%%% 58
\bibitem[\protect\citeauthoryear{{Statistics Canada}}{2016b}]{hh_pumf}
\begin{botherref}
\oauthor{\bsnm{{Statistics Canada}}}:
{Hierarchical File, 2016 Census of Population – Catalogue no. 98M0002X}.
Available at \url{https://www150.statcan.gc.ca/n1/en/catalogue/98M0002X}.
Accessed: 2022-08
(2016)
\end{botherref}
\endbibitem

%%% 59
\bibitem[\protect\citeauthoryear{{Statistics Canada}}{2016c}]{indiv_pumf}
\begin{botherref}
\oauthor{\bsnm{{Statistics Canada}}}:
{Individuals File, 2016 Census of Population – Catalogue no. 98M0001X}.
Available at \url{https://www150.statcan.gc.ca/n1/en/catalogue/98M0001X}.
Accessed: 2022-08
(2016)
\end{botherref}
\endbibitem

%%% 60
\bibitem[\protect\citeauthoryear{{Statistics Canada}}{2017}]{canada_senior_worker}
\begin{botherref}
\oauthor{\bsnm{{Statistics Canada}}}:
Working seniors in canada.
Technical report,
{Statistics Canada}
(2017).
\url{https://www12.statcan.gc.ca/census-recensement/2016/as-sa/98-200-x/2016027/98-200-x2016027-eng.cfm}
\end{botherref}
\endbibitem

%%% 61
\bibitem[\protect\citeauthoryear{{Statistics Canada}}{2018}]{proj_2018}
\begin{botherref}
\oauthor{\bsnm{{Statistics Canada}}}:
{Projected population, by projection scenario, age and sex, as of July 1 (x 1,000) – Table 17-10-0057-01}.
Available at \url{https://www150.statcan.gc.ca/t1/tbl1/en/tv.action?pid=1710005701}.
Accessed: 2022-08
(2018)
\end{botherref}
\endbibitem

%%% 62
\bibitem[\protect\citeauthoryear{{Statistics Canada}}{2022a}]{census_2021}
\begin{botherref}
\oauthor{\bsnm{{Statistics Canada}}}:
{Census Profile. 2021 Census of Population. Statistics Canada Catalogue number 98-316-X2021001}.
\url{https://www12.statcan.gc.ca/census-recensement/2021/dp-pd/prof/index.cfm?Lang=E}
(2022)
\end{botherref}
\endbibitem

%%% 63
\bibitem[\protect\citeauthoryear{{Statistics Canada}}{2022b}]{health_survey}
\begin{botherref}
\oauthor{\bsnm{{Statistics Canada}}}:
{Health characteristics, annual estimates – Table 13-10-0096-01}.
Available at \url{https://doi.org/10.25318/1310009601-eng}.
Accessed: 2022-08
(2022).
\doiurl{10.25318/1310009601-eng}
\end{botherref}
\endbibitem

%%% 64
\bibitem[\protect\citeauthoryear{{StreetLight Data}}{2021}]{aia_2021}
\begin{botherref}
\oauthor{\bsnm{{StreetLight Data}}}:
Quarterly report: National vehicle kilometres traveled metrics – 2019 to 2021.
Technical report,
AIA Canada
(2021).
\url{https://www.aiacanada.com/product/national-vehicle-kilometres-traveled-metrics-from-2019-to-2021-q1-and-q2-results/}
\end{botherref}
\endbibitem

%%% 65
\bibitem[\protect\citeauthoryear{Shergold and Wilson}{2016}]{shergold_mobility_2016}
\begin{botherref}
\oauthor{\bsnm{Shergold}, \binits{I.}},
\oauthor{\bsnm{Wilson}, \binits{M.}}:
The mobility of older people, and the future role of connnected autonomous vehicles.
Technical report,
Centre for Transport \& society, University of the West of England
(2016)
\end{botherref}
\endbibitem

%%% 66
\bibitem[\protect\citeauthoryear{{United Nations, Department of Economic and Social Affairs, Population Division}}{2019}]{un_2019}
\begin{botherref}
\oauthor{\bsnm{{United Nations, Department of Economic and Social Affairs, Population Division}}}:
World Population Ageing 2019: Highlights (ST/ESA/SER.A/430)
(2019)
\end{botherref}
\endbibitem

%%% 67
\bibitem[\protect\citeauthoryear{Viergutz and Schmidt}{2019}]{viergutz_demand_2019}
\begin{barticle}
\bauthor{\bsnm{Viergutz}, \binits{K.}},
\bauthor{\bsnm{Schmidt}, \binits{C.}}:
\batitle{Demand responsive - vs. conventional public transportation: A {MATSim} study about the rural town of {Colditz, Germany}}.
\bjtitle{Procedia Computer Science}
\bvolume{151},
\bfpage{69}--\blpage{76}
(\byear{2019})
\doiurl{10.1016/j.procs.2019.04.013}
\end{barticle}
\endbibitem

%%% 68
\bibitem[\protect\citeauthoryear{Wadud et~al.}{2016}]{wadud_2016}
\begin{botherref}
\oauthor{\bsnm{Wadud}, \binits{Z.}},
\oauthor{\bsnm{Don}, \binits{M.}},
\oauthor{},
\oauthor{\bsnm{Paul}, \binits{L.}}:
Help or hindrance? the travel, energy and carbon impacts of highly automated vehicles.
Transportation Research Part A: Policy and Practice,
1--18
(2016)
\doiurl{10.1016/j.tra.2015.12.001}
\end{botherref}
\endbibitem

%%% 69
\bibitem[\protect\citeauthoryear{}{}]{winnipeg_tomtom}
\begin{botherref}
Winnipeg traffic report {\textbar} {TomTom} Traffic Index.
\url{https://www.tomtom.com/traffic-index/winnipeg-traffic/}
Accessed 2023-07-12
\end{botherref}
\endbibitem

%%% 70
\bibitem[\protect\citeauthoryear{{Winnipeg open data portal}}{2022}]{addresses}
\begin{botherref}
\oauthor{\bsnm{{Winnipeg open data portal}}}:
{City of Winnipeg currently active official property addresses and associated coordinates}.
Available at \url{https://data.winnipeg.ca/City-Planning/Addresses/cam2-ii3u}.
Accessed: 2022-05
(2022)
\end{botherref}
\endbibitem

%%% 71
\bibitem[\protect\citeauthoryear{Wang et~al.}{2018}]{wang_simulation_2018}
\begin{barticle}
\bauthor{\bsnm{Wang}, \binits{B.}},
\bauthor{\bsnm{Medina}, \binits{S.A.O.}},
\bauthor{\bsnm{Fourie}, \binits{P.}}:
\batitle{Simulation of autonomous transit on demand for fleet size and deployment strategy optimization}.
\bjtitle{Procedia Computer Science}
\bvolume{130},
\bfpage{797}--\blpage{802}
(\byear{2018})
\doiurl{10.1016/j.procs.2018.04.138}
\end{barticle}
\endbibitem

%%% 72
\bibitem[\protect\citeauthoryear{Watson et~al.}{2021}]{watson_2021}
\begin{botherref}
\oauthor{\bsnm{Watson}, \binits{K.B.}},
\oauthor{\bsnm{Whitfield}, \binits{G.P.}},
\oauthor{\bsnm{Bricka}, \binits{S.}},
\oauthor{\bsnm{Carlson}, \binits{S.A.}}:
Purpose-based walking trips by duration, distance, and select characteristics, 2017 national household travel survey.
Journal of physical activity \& health,
86--93
(2021)
\doiurl{10.1123/jpah.2021-0096}
\end{botherref}
\endbibitem

%%% 73
\bibitem[\protect\citeauthoryear{Zandieh and Acheampong}{2021}]{zandieh2021mobility}
\begin{barticle}
\bauthor{\bsnm{Zandieh}, \binits{R.}},
\bauthor{\bsnm{Acheampong}, \binits{R.A.}}:
\batitle{Mobility and healthy ageing in the city: Exploring opportunities and challenges of autonomous vehicles for older adults' outdoor mobility}.
\bjtitle{Cities}
\bvolume{112},
\bfpage{103135}
(\byear{2021})
\end{barticle}
\endbibitem

%%% 74
\bibitem[\protect\citeauthoryear{Ziemke et~al.}{2019}]{ziemke_2019}
\begin{bchapter}
\bauthor{\bsnm{Ziemke}, \binits{D.}},
\bauthor{\bsnm{Kaddoura}, \binits{I.}},
\bauthor{\bsnm{Nagel}, \binits{K.}}:
\bctitle{{The MATSim open Berlin scenario: A multimodal agent-based transport simulation scenario based on synthetic demand modeling and open data}}.
In: \bbtitle{Procedia Computer Science},
vol. \bseriesno{151},
pp. \bfpage{870}--\blpage{877}
(\byear{2019})
\end{bchapter}
\endbibitem

\end{thebibliography}

\end{document}